\documentstyle[12pt]{article}

\input epsf
\makeatletter

\def\newpic#1{}

\voffset= - 0.5cm

\def\hybrid{\topmargin 0pt      \oddsidemargin 0pt
             \headheight 0pt \headsep 0pt

       \textwidth 6.5in        
       \textheight 9in         
             \marginparwidth 0.0in
             \parskip 5pt plus 1pt   \jot = 1.5ex}
\catcode`\@=11
\def\marginnote#1{}

\newcount\hour
\newcount\minute
\newtoks\amorpm
\hour=\time\divide\hour by60
\minute=\time{\multiply\hour by60 \global\advance\minute by-\hour}
\edef\standardtime{{\ifnum\hour<12 \global\amorpm={am}%
             \else\global\amorpm={pm}\advance\hour by-12 \fi
             \ifnum\hour=0 \hour=12 \fi
             \number\hour:\ifnum\minute<10 0\fi\number\minute\the\amorpm}}
\edef\militarytime{\number\hour:\ifnum\minute<10 0\fi\number\minute}

\def\draftlabel#1{{\@bsphack\if@filesw {\let\thepage\relax
        \xdef\@gtempa{\write\@auxout{\string
           \newlabel{#1}{{\@currentlabel}{\thepage}}}}}\@gtempa
        \if@nobreak \ifvmode\nobreak\fi\fi\fi\@esphack}
             \gdef\@eqnlabel{#1}}
\def\@eqnlabel{}
\def\@vacuum{}
\def\draftmarginnote#1{\marginpar{\raggedright\scriptsize\tt#1}}

\def\draftlabel#1{{\@bsphack\if@filesw {\let\thepage\relax
        \xdef\@gtempa{\write\@auxout{\string
           \newlabel{#1}{{\@currentlabel}{\thepage}}}}}\@gtempa
        \if@nobreak \ifvmode\nobreak\fi\fi\fi\@esphack}
             \gdef\@eqnlabel{#1}}
\def\@eqnlabel{}
\def\@vacuum{}
\def\draftmarginnote#1{\marginpar{\raggedright\scriptsize\tt#1}}

\def\draft{\oddsidemargin -.5truein
             \def\@oddfoot{\sl preliminary draft \hfil
             \rm\thepage\hfil\sl\today\quad\militarytime}
             \let\@evenfoot\@oddfoot \overfullrule 3pt
             \let\label=\draftlabel
             \let\marginnote=\draftmarginnote
        \def\@eqnnum{(\theequation)\rlap{\kern\marginparsep\tt\@eqnlabel}%
\global\let\@eqnlabel\@vacuum}  }

\def\numberbysection{\@addtoreset{equation}{section}
             \def\theequation{\thesection.\arabic{equation}}}

\def\underline#1{\relax\ifmmode\@@underline#1\else
             $\@@underline{\hbox{#1}}$\relax\fi}

\def\titlepage{\@restonecolfalse\if@twocolumn\@restonecoltrue\onecolumn
          \else \newpage \fi \thispagestyle{empty}\c@page\z@
             \def\thefootnote{\fnsymbol{footnote}} }

\def\endtitlepage{\if@restonecol\twocolumn \else  \fi
             \def\thefootnote{\arabic{footnote}}
             \setcounter{footnote}{0}}  
\relax

\makeatletter
\newdimen\normalarrayskip              
\newdimen\minarrayskip                 
\normalarrayskip\baselineskip
\minarrayskip\jot
\newif\ifold             \oldtrue            \def\new{\oldfalse}
\def\arraymode{\ifold\relax\else\displaystyle\fi} 
\def\eqnumphantom{\phantom{(\theequation)}}     
\def\@arrayskip{\ifold\baselineskip\z@\lineskip\z@
         \else
         \baselineskip\minarrayskip\lineskip2\minarrayskip\fi}
\def\@arrayclassz{\ifcase \@lastchclass \@acolampacol \or
\@ampacol \or \or \or \@addamp \or
       \@acolampacol \or \@firstampfalse \@acol \fi
\edef\@preamble{\@preamble
      \ifcase \@chnum
         \hfil$\relax\arraymode\@sharp$\hfil
         \or $\relax\arraymode\@sharp$\hfil
         \or \hfil$\relax\arraymode\@sharp$\fi}}
\def\@array[#1]#2{\setbox\@arstrutbox=\hbox{\vrule
         height\arraystretch \ht\strutbox
         depth\arraystretch \dp\strutbox
         width\z@}\@mkpream{#2}\edef\@preamble{\halign
\noexpand\@halignto
\bgroup \tabskip\z@ \@arstrut \@preamble \tabskip\z@ \cr}%
\let\@startpbox\@@startpbox \let\@endpbox\@@endpbox
      \if #1t\vtop \else \if#1b\vbox \else \vcenter \fi\fi
      \bgroup \let\par\relax
      \let\@sharp##\let\protect\relax
      \@arrayskip\@preamble}
\def\eqnarray{\stepcounter{equation}%
                  \let\@currentlabel=\theequation
                  \global\@eqnswtrue
                  \global\@eqcnt\z@
                  \tabskip\@centering
                  \let\\=\@eqncr
     \halign to \displaywidth\bgroup
        \eqnumphantom\@eqnsel\hskip\@centering
        $\displaystyle \tabskip\z@ {##}$%
        \global\@eqcnt\@ne \hskip 2\arraycolsep
             $\displaystyle\arraymode{##}$\hfil
        \global\@eqcnt\tw@ \hskip 2\arraycolsep
             $\displaystyle\tabskip\z@{##}$\hfil
             \tabskip\@centering
        &{##}\tabskip\z@\cr}
\begingroup\ifx\undefined\newsymbol \else\def\input#1 {\endgroup}\fi
\newfont{\hr}{msbm10}
\newfont{\ams}{msam10}

\def\beq{\begin{equation}}
\def\eeq{\end{equation}}
\def\ba{\beq\new\begin{array}{c}}
\def\ea{\end{array}\eeq}

\def\p{\partial}

\def\Doil{{\sf D_{\rm oil}}}
\def\Dalpha{{\sf D}_{\alpha}}
\def\Dbeta{{\sf D}_{\beta}}
\def\D1{{\sf D}_{1}}
\def\Dg{{\sf D}_{g}}
\def\aalpha{{\sf a}_{\alpha}}
\def\balpha{{\sf b}_{\alpha}}

\hybrid

\begin{document}

\begin{titlepage}

\title{Laplacian growth
and Whitham equations of soliton theory}

\author{I.~Krichever \thanks{Department of Mathematics, Columbia
University, New York, USA, Landau Institute and ITEP, Moscow, Russia}
\and M.~Mineev-Weinstein
\thanks{ Los Alamos National Laboratory, MS-P365, Los Alamos, NM 87545, USA }
\and
P.~Wiegmann \thanks{James Frank Institute and Enrico Fermi
Institute
of the University of Chicago, 5640 S.Ellis Avenue,
Chicago, IL 60637, USA and
Landau Institute for Theoretical Physics, Moscow, Russia}
\and A.~Zabrodin
\thanks{CNLS, LANL, Los Alamos, NM 87545, USA,
and Institute of Biochemical Physics,
4 Kosygina st., 119991, Moscow, Russia
and ITEP, 25 B.Cheremushkinskaya, 117259,
Moscow, Russia}}

\date{October 2003}
\maketitle

\begin{abstract}

The Laplacian growth (the Hele-Shaw problem) of multiply-connected
domains in the case of zero
surface tension
is proven to be equivalent to an integrable system
of Whitham equations known in soliton theory. The Whitham equations
describe slowly modulated periodic solutions  of integrable hierarchies
of nonlinear differential equations.
  Through this connection the
Laplacian growth is understood as a flow in the moduli space of
  Riemann surfaces.
\end{abstract}

\vfill

\end{titlepage}

\section{Introduction}
It is not uncommon that nonlinear differential equations which
possess an integrable structure emerge in important problems of
hydrodynamics \cite{Novikov}. The Korteweg de Vries equation describing
nonlinear waves in dispersive media is perhaps the most familiar example.

In recent
years, integrable structures were found in another class of
hydrodynamics problems leading to a pattern formation in a regime far
from equilibrium \cite{MWZ}.
Growth problems of this type are unified by the name
Laplacian growth. In this paper, we further develop a link between growth
processes and soliton theory. We extend the results of Ref.\cite{MWZ} to
the case of multiply-connected domains and identify
the set of growth processes with a universal
Whitham hierarchy of integrable equations. The latter unveils the
mathematical structure of the growth and set a place for growth models
in the realm of soliton theory.

Laplacian growth, also known as the Hele-Shaw problem, refers to
dynamics of a moving front (an interface) between two distinct phases
driven by a harmonic scalar field.  This field is a  potential  for the
growth velocity field.
The Laplacian growth problem
appears in different physical and mathematical contexts
and has a number of  important practical applications.
The most known ones are filtration processes in porous media,  viscous
fingering in the Hele-Shaw cell, electrodeposition and solidification in
undercooled liquids.   A comprehensive list of relevant papers
published prior to 1998  can be found in
\cite{list}.

\begin{figure}[tb]
\epsfysize=3.7cm
\centerline{\epsfbox{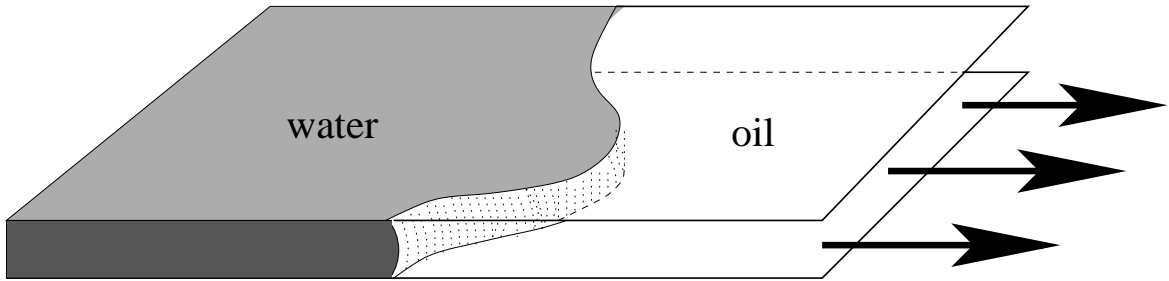}}
\caption{\sl The Hele-Shaw cell (arrows show direction of the
flow forced by a pump). }
\label{fi:heleshaw}
\end{figure}

The most interesting and the most studied dynamics occurs in
two-dimensional spatial geometry.
To be definite, we shall speak about an interface between two
incompressible fluids with very different viscosities on the plane.
In practice the 2D geometry is
realized in a Hele-Shaw cell - a narrow gap between two parallel
plates (Fig.~\ref{fi:heleshaw}).
In this version, the problem is also  known as the Saffman-Taylor problem
or viscous fingering. For a review, see \cite{RMP}.
Importance of studies of the Laplacian growth
with more than one nonviscous droplet speaks for itself.  When rates of flow
are
considerable, then, because of fingering instability,
new droplets are pinched
off and change their shapes, so the whole dynamics considerably changes
in comparison with a single bubble dynamics.
(See, e.g., experimental works \cite{SwinneyMaxworthy}.)

To be more precise, consider the case when there are several
disconnected domains in the Hele-Shaw cell occupied by
a fluid with low viscosity (water).
We call them water droplets.
Their exterior, which is in general a multiply-connected domain,
is occupied by a viscous fluid (oil).
All components
of the oil/water interface are assumed to be smooth curves.
Oil is sucked out with fixed rates $Q_j$ through sinks placed
at some finite points $a_j$ or at infinity (edges of the Hele-Shaw cell).
Water is injected
into each water droplet with rates
$q_{\alpha}$, some of which
may be negative or equal to zero.
Hele-Shaw flows in
general setting, including the multiply-connected
case, were discussed in \cite{Rich}-\cite{EV}.

Viscous flows are governed by gradient of the pressure
field in the fluids.
In the oil domain, the local velocity
$\vec V=(V_x,V_y)$ of the fluid is proportional to
the gradient of pressure $p=p(x,y)$
(Darcy's law):
$\vec V=-\kappa \nabla p$, where $\kappa$ is called
the filtration coefficient. In what follows, we choose units
in such a way that $\kappa =\frac{1}{4}$.
In particular, the Darcy law holds on the
outer side of the interface thus
governing its dynamics:
\beq
\label{lg1a}
V_n=-\frac{1}{4} \p_n p\,.
\eeq
Here $\p_n $ is the normal derivative.

This simple dynamics results to complicated unstable patterns often
growing beyond control. The most recent experimentally produced pattern
can be seen in Ref. \cite{SwinneyMaxworthy}.

In this paper we discuss only the idealized
problem, namely that surface tension equals zero.  We call it
{\it the idealized Laplacian growth (ILG)}.
Zero surface tension means that pressure
does not jump across the boundary.
Assuming that
viscosity of water is small enough comparing
to the viscosity of oil, pressure is constant inside each
water droplet.
However, the values of pressure
may be different in different droplets
and may also depend on time.
Let $p_{\alpha}$ be pressure
in the $\alpha$-th droplet, then zero surface tension
implies that
$p=p_{\alpha}$ on the outer side of the interface as well.

Since fluids are incompressible ($\nabla \vec V =0$)
the Darcy law implies
that the pressure field $p$
is a harmonic function in the exterior (oil) domain
except at the points where the oil pumps are located.
In the case of zero surface tension, pressure is a solution
of the time-dependent boundary problem for the Laplace
equation with $p=p_{\alpha}$ on the boundary components.
The interface moves according to
the Darcy law (\ref{lg1a}), so that the boundary problem
changes with time. Note
that the problem is non-local since the gradient of pressure around
boundary depends on the shape of the domain as a whole.

When the interface bounds
a simply-connected domain, an effective tool
for dealing with the
Laplacian growth is the time dependent conformal
mapping technique (see e.g. \cite{RMP}).
Passing to the complex coordinates $z=x+iy$,
$\bar z =x-iy$ on the physical plane,
one may describe dynamics in terms of
a moving conformal map from
a simple reference domain,
say the unit disk in a ``mathematical plane'',
onto a growing domain in the physical plane.

If the interface has several
disconnected components the conformal map
approach meets fundamental difficulties. Uniformizing
maps of multiply-connected domains are essentially more complicated
mainly because there are no simple
reference domains and, moreover, any
possible reference domain depends on the
growing domain itself.

Instead, given growing domains, one may construct an
algebraic curve over complex numbers, or
a Riemann surface,
endowed with some additional structures. Then
the Darcy law  translates the growth  to an evolution of  the
algebraic curve, thus providing a remarkable family
of flows on the (extended) moduli
space of Riemann surfaces.

This is  the meeting point of the growth processes and soliton theory.
We show that the evolution of the Riemann surface  built upon the moving
interface in the Hele-Shaw cell
is governed by the universal Whitham hierarchy of soliton
theory. It generalizes the
dispersionless 2D Toda hierarchy which was shown
in Ref. \cite{MWZ} to describe the moving boundary problems
in the simply-connected case.
The Whitham equations have been originally introduced
to describe slow
modulations of periodic solutions to differential equations
\cite{Whitham}. Their relation to the
algebraic geometry, using example of the KdV equation, was found in
\cite{FFM}. In general
setting, the universal Whitham hierarchy was introduced in
\cite{kri1,kri2,kri3}.

The mathematical equivalence
between the Laplacian growth with zero surface tension
and the Whitham equations, established in this work,
allows one to treat the former as
a singular limit of dispersive waves obeying soliton
equations. Such a singular limit may often lead to
solutions which develop singularities
within finite time.

In the context of the Laplacian
growth, the singularities show up as cusps
generated by an initially
smooth interface \cite{SB}, after which
the idealized description no longer makes
any physical sense.
This feature signifies that
the very problem with zero surface tension is ill-posed.

A similar difficulty has been known to exist
for the Whitham equations (see, e.g., \cite{Singular}).
Some of their solutions, being initially regular,
sooner or later become singular and so
can not be extended to all times.
The Korteweg de Vries equation with
zero dispersion (the Hopf equation) is the most familiar example.
Here any smooth decreasing
Cauchy data evolve into a ``shock wave" with an overturned
front which is physically meaningless (a ``gradient catastrophe").
This simply means that the dispersionless approximation
does not work in a vicinity of the catastrophe.
Similarly, the
zero surface tension assumption is not valid
in a  vicinity of the cusp formation.

These singularities are in fact artificial and
can be successfully resolved by methods
developed in the theory of
slow modulations of exact periodic solutions to
soliton equations \cite{Novikov,Regularizing}.
In subsequent works we hope to apply these methods to the
Laplacian growth using the proven
below equivalence between the two disciplines.

\section{Linearization of the ILG dynamics}

Remarkably, the ILG dynamics, initially formulated as
a non-local and highly nonlinear problem, admits
an exact {\it linearization} in the space of harmonic moments
of the viscous domain.
By linearization we mean here a change
of variables which converts the nontrivial LG dynamics
into a simple linear one. A familiar (but rather loose)
analogy is passing to action-angle variables
in classical mechanics or the inverse scattering
transform in the soliton theory.

Our starting point is the fact that
the ILG
is a simple linear shift
in the space of harmonic moments
of the growing domain.
This statement goes back to
seminal Richardson's paper \cite{Rich}.
In that paper, it
was shown that if there is the only sink at
infinity, all moments are conserved except the moment
of constant function
(the area of the droplet) which changes linearly with time.
In fact, it is absolutely
clear that for incompressible
fluids and fixed pumping rates
areas of the droplets, if change at all,
always do this linearly with time.
For arbitrary location of the sink, and also for
several sinks at different points, a simple extension
of this result states that in general all moments
change linearly with time (with different coefficients
which may be zero).

In the case of several
water droplets, the set of harmonic moments
should be supplemented by
a finite number of extra parameters, one for each
extra droplet, which are
basically moments of harmonic functions with
multivalued analytic parts.
This set of variables is enough
to characterize the geometry of the growing domain.
Alternatively,
the new parameters may be areas of the water droplets.
Depending on which type of external physical conditions
in the water droplets is realized (fixed pressure differences or
fixed pumping rates), the ILG dynamics becomes linear
either in the former or in the latter variables.

Most of the material of Secs. 2.1 and 2.2 is spread through the
literature (see, e.g.,
\cite{Rich,Rich1,Rich2,Etingof} and Chapter 5 of the book
\cite{EV}). To make our exposition self-contained,
we review them from a unifying
point of view.

\subsection{The time-dependent boundary problem}

Consider an ILG process with the point-like oil pumps
with powers $Q_j$ at some points $a_j$
located far enough from the moving interface.
Mathematically this means
\beq\label{caj}
-\oint_{{\sf c}_{j}} V_n ds =\pi Q_j\,,
\eeq
where ${\sf c}_{j}$ is a small contour encircling the point
$a_j$, $ds$ stands for
the differential of the arc length,
$V_n$ is the component of the fluid velocity normal to
the contour,
with the normal vector pointing
outside the circle ${\sf c}_{j}$.
Both $Q_j$ and $a_j$ are assumed to be
time-independent.
Oil may be also sucked at infinity. Physically this
  means, for example, that oil is removed from the edge of
a large Hele-Shaw cell. Mathematically one puts
one of $a_j$ equal to infinity and defines the
pumping rate at infinity, $Q_{\infty}$, as
$$
-\oint_{{\sf c}_{\infty}} V_n ds =\pi Q_{\infty}\,,
$$
Here ${\sf c}_{\infty}$ is a big contour encircling
the whole system of water droplets and all the point-like
pumps, if any. The oil pumping rates are assumed
to be positive when oil is sucked and negative
if it is injected into the Hele-Shaw cell.

One can also consider
extended sources or sinks of oil,
for instance, continuously distributed
along lines, like in Ref. \cite{Etingof}.
To avoid irrelevant technical complications, we
consider point-like oil pumps only, giving
brief remarks on the more general case when necessary.

\begin{figure}[tb]
\epsfysize=7cm
\centerline{\epsfbox{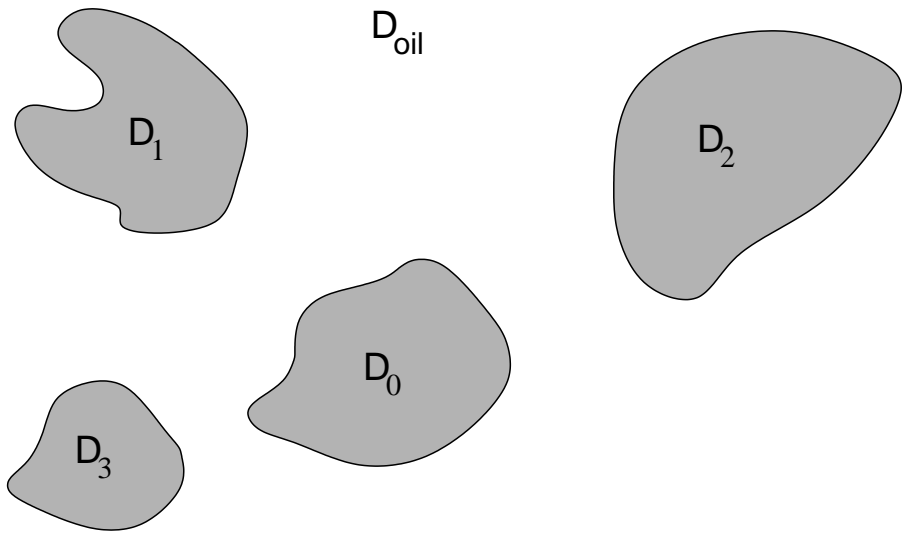}}
\caption{\sl Water droplets $\Dalpha$ ($g=3$).}
\label{fi:droplets}
\end{figure}

Let $\Doil$ be the region of the plane occupied by oil (an infinite
domain containing infinity), and ${\sf D}$ be the region occupied by
water.
We assume that there are $g+1$ water droplets
in the Hele-Shaw cell, which are compact domains
bounded by smooth non-intersecting curves.
Let $\Dalpha $ be the $\alpha$-th water droplet,
$\alpha =0,1, \ldots , g$, so that ${\sf D}$
is their union (Fig.~\ref{fi:droplets}).
It is convenient to think of the $0$-th droplet
as of the ``main'' one, having in mind that it is this droplet
that remains in the system in the simply-connected case.
Let us assume, just for a notational convenience,
that the origin is in the zero
droplet, $0\in {\sf D}_0$.

The incompressibility implies that the oil pumps
are able to work only if there are some sources of water
which supply water to at least one of the droplets.
We call them water pumps and introduce
the pumping rate $q_{\alpha}$ of water in the $\alpha$-th
droplet:
\beq\label{qq}
\oint_{\p \Dalpha } V_n ds =\pi q_{\alpha}\,,
\eeq
Here the normal vector points
{\it outside} $\Dalpha$. The pumping rate $q_{\alpha}$
is assumed to be positive when water is injected into the
$\alpha$-droplet. Obviously, the oil and water pumping rates
are constrained by the relation
$$
Q_{\infty}+\sum_j Q_j =\sum_{\alpha}q_{\alpha}
$$

The pressure field in oil obeys the equation
\beq
\label{lg2}
\Delta p(z) =4\pi \sum_j Q_j \delta^{(2)}(z-a_j )\,,
\eeq
where
$\delta^{(2)}(z)$ is the two-dimensional delta-function and
$\Delta =\p_{x}^{2}+\p_{y}^{2} =4\p_z \p_{\bar z}$
is the Laplace operator on the plane.
In other words, $p =p(z)$ is a harmonic function in $\Doil$
with the asymptote
$$
p(z)=Q_j \log |z-a_j |^2 +\ldots \;\;\;
\mbox{in the vicinity of $a_j$}
$$
Taking into account the Darcy law, this agrees
with the definition of the $Q_j$ (\ref{caj}).
On the $\alpha$-th component
of the interface the pressure
field takes a constant value $p_{\alpha}$
(which may depend on time).
Without loss of generality, we set $p_0 =0$.

To determine
velocity of the interface, one
should solve a time-dependent
Dirichlet-like boundary problem.
It has a unique solution, and so pressure
in $\Doil$ is uniquely determined as soon as
one specifies pressures
$p_{\alpha}$ in water droplets. This means that
with given $p_{\alpha}$'s there is no freedom
in water pumping. Indeed, the pumping rates in
all water droplets are to be determined from
(\ref{qq}), which states that
$q_{\alpha}=-\frac{1}{4\pi}\oint_{\p \Dalpha}
\p_n p \, ds$. If $p_{\alpha}$'s are maintained
constant with time, $q_{\alpha}$'s are in general certain
complicated functions of time. Alternatively,
one may control the pumping rates $q_{\alpha}$
keeping them constant, then pressures in
water droplets are uniquely determined by the
dynamical equations and in general exhibit
a nontrivial time dependence.

So the physical problem is not yet defined by the
local growth law alone. One should add
some physical conditions in water droplets.
We  distinguish two cases:
\begin{itemize}
\item [(I)]
Fixed pressure differences $p_{\alpha}$:
$\dot p_{\alpha} =0$, then $q_{\alpha} =q_{\alpha}(t)$ are to be
determined;
\item [(II)]
Fixed pumping rates $q_{\alpha}$:
  $\dot q_{\alpha} =0$, then $p_{\alpha}=p_{\alpha}(t)$ are to be
determined
\end{itemize}
Here the dot means the time derivative.
Various ``mixed'' conditions (say, when we fix
some of $p_{\alpha}$'s and some of $q_{\alpha}$'s or
linear combinations thereof) are not meaningless
from the mathematical point of view but look somewhat
artificial in the physical context.

For an abuse of quantum-mechanical terminology, one may
refer to the cases (I) and (II) as the LG problem in
$p$-representation and $q$-representation, respectively.

\subsection{Integral transforms of the oil domain}

\paragraph{The Cauchy transform.}
Here we closely follow Richardson's paper \cite{Rich2}.
Let us consider the Cauchy transform of the oil domain:
\beq\label{Cauchy1}
h(z)=\frac{1}{\pi}\int_{\Doil}\frac{d^2 \zeta}{z-\zeta}
\eeq
(A cut-off at some radius $R$ is implied,
at large $R$  the integral does not depend
on the cut-off which may be safely moved to infinity.)
This function is continuous across
all boundaries and
analytic for $z$ in each hole $\Dalpha$
while in $\Doil$ the function $h(z)-\bar z$ is analytic.
So we can write
\beq\label{Cauchy2}
h(z)=\left \{
\begin{array}{l}
h_{\alpha}^{+}(z) \;\;\;\; \mbox{for $z$ in $\Dalpha$}
\\ \\
\bar z + h^-(z) \;\; \mbox{for $z$ in $\Doil$}
\end{array}
\right.
\eeq
where $h_{\alpha}^{+}(z)$ is an analytic function in
$\Dalpha$ and $h^-(z)$ is analytic for $z$ in
$\Doil$.
One may analytically continue
the functions $h_{\alpha}^{+}$, $h^-$ outside the
regions where they are defined by the Cauchy transform.
In general, all $g+1$ functions $h_{\alpha}^{+}(z)$ obtained
in this way are expected to be different.

The contour integral representation of these functions
reads:
\beq\label{Cauchy3}
\frac{1}{2\pi i}\oint_{\p {\sf D}}
\frac{\bar \zeta d\zeta}{\zeta -z} =
\left \{
\begin{array}{l}
h_{\alpha}^{+}(z) \;\;\;\; \mbox{for $z$ in $\Dalpha$}
\\ \\
h^-(z) \;\; \mbox{for $z$ in $\Doil$}
\end{array}
\right.
\eeq
Note that $h_{\alpha}^{+}(z)-h^-(z)=\bar z$ on the
$\alpha$-th boundary curve.

The time derivative
$\dot h (z) =\p_t h(z;t)$, is found
straightforwardly using the integral representation
(\ref{Cauchy1}) and the Darcy law:
$$
\dot h(z)=
-\frac{1}{\pi}\oint_{\p \Doil (t)}
\frac{1}{z-\zeta}\,
V_n ds =
\frac{1}{4\pi} \oint_{\p \Doil (t)}
\frac{1}{z-\zeta}\,
\p_n p (\zeta )ds =
$$
$$
=\frac{1}{4\pi}\oint_{\p \Doil (t)}
\left (
\frac{1}{z-\zeta}\,
\p_n p (\zeta ) -p(\zeta ) \p_n \, \frac{1}{z-\zeta}\,
\right ) ds
+\frac{1}{4\pi}\sum_{\alpha =1}^{g} p_{\alpha}
\oint_{\p \Dalpha (t)} \p_n \, \frac{1}{z-\zeta}\, ds
$$
where we subtracted and added the integral of
$p \p_n \frac{1}{z-\zeta}$ over the whole boundary, and used
the fact that $p$ is constant along any component of the
boundary. It is easy to see that irrespectively of whether
the point $z$ is inside or outside $\Dalpha$, all contour
integrals in the second term vanish.
The first term can be transformed using the Green theorem:
$$
\oint_{\p \Doil (t)}
\left (\frac{1}{z-\zeta}\, \p_n p -p
\p_n \frac{1}{z-\zeta}\right ) ds =
-\int_{ \Doil (t)}
\left (\frac{1}{z-\zeta}\, \Delta p (\zeta) -p(\zeta)
\Delta \, \frac{1}{z-\zeta}\right ) d^2 \zeta
$$
The Laplacian of $p(\zeta)$ is given by (\ref{lg2}).
If $z$ is inside a water droplet,
the function $\frac{1}{z-\zeta}$ (regarded as a function
of $\zeta$) is harmonic for all $\zeta$ in $\Doil$
and the second term vanishes.
If $z$ is in $\Doil$,
$$
\int_{ \Doil (t)}
p(\zeta) \Delta_{\zeta}
\left (\frac{1}{z-\zeta}\right ) d^2 \zeta =
4\pi \p_z p(z)
$$
Finally, we get:
\beq\label{hdot}
\dot h_{\alpha}^{+}(z)=\sum_j \frac{Q_j}{a_j -z}
\;\;\;\;
\mbox{for $z$ in $\Dalpha$}
\eeq
\beq\label{htildedot}
\dot h^-(z)=\sum_j \frac{Q_j}{a_j -z}
+\p_z p(z)
\;\;\;\;
\mbox{for $z$ in $\Doil$}
\eeq
In the integrated form,
\beq\label{ht}
h_{\alpha}^{+}(z;t)=
h_{\alpha}^{+}(z;0)+t
\sum_j \frac{Q_j}{a_j -z}
\eeq
We see that the increment of
the $h_{\alpha}^{+}$ is a rational function,
and, moreover, it is {\it the same} for all $\alpha =
0,1, \ldots , g$. We also see that this function is
entirely determined by the output powers and locations
of oil pumps, no matter what conditions we impose on the
pressures and pumping rates in water droplets.

For a
linearly distributed oil source,  $p(z)$ is
  a  potential of a simple layer. Then  one finds that
the functions $\dot h_{\alpha}^{+}$, though no longer rational,
are still analytic continuations of a single
analytic function
unless the support of the simple layer forms a noncontractable
cycle encircling at least one of the water droplets.

A remarkable property of the Cauchy transform
is its  linear dependence on time $t$ (provided the
parameters of the oil pumps are time-independent).
This is the key point that allows one to linearize
the ILG dynamics.

\paragraph{The Coulomb potential.}

Along with the Cauchy transform of the domain $\Doil$
it is useful to consider the potential generated by
fictitious 2D Coulomb charges uniformly distributed
in $\Doil$:
\beq\label{pi1}
\phi (z)=\frac{1}{\pi} \int_{\Doil}
\log \left | 1-\frac{z}{\zeta}\right |^2 d^2 \zeta
\eeq
(The same cut-off as in the Cauchy
transform is implied.)
Clearly, this function is
harmonic in each water droplet
and its $z$-derivative coincides
with the $h(z)$: $\p_z \phi (z) =h(z)$.
Repeating the above calculation for $\phi (z)$
we get (for $z$ in water droplets):
$$
\dot \phi (z)=
-\frac{1}{4\pi} \int_{\Doil}
\log \left | 1-\frac{z}{\zeta}\right |^2
\Delta p (\zeta ) d^2 \zeta
+\frac{1}{4\pi} \sum_{\alpha =1}^{g}
p_{\alpha}
\oint_{\p \Dalpha} \p_n
\log \left | 1-\frac{z}{\zeta}\right |^2
ds
$$
This yields:
\beq\label{phidot}
\dot \phi (z)=
p_{\alpha} - \sum_j Q_j
\log \left | 1-\frac{z}{a_j}\right |^2
\;\;\;\;\;
\mbox{for $z$ in $\Dalpha$}
\eeq

\paragraph{The Laplacian growth equation.}
For completeness, let us demonstrate that
the Laplacian growth equation,
usually derived, in the simply-connected case,
using the
time dependent conformal map technique,
follows from the time derivative of the
Cauchy transform.

To this end, we calculate the difference of the
boundary values $\dot h^{\pm}$ in two ways.
On the one hand, it is
obtained by subtracting (\ref{hdot}) and
(\ref{htildedot}):
\beq\label{onehand}
\dot h^{+}(z)-\dot h^{-}(z) =-\p_z p(z)\,,
\;\;\;\;\mbox{for $z$ on any boundary contour}
\eeq
On the other hand,
$$
\p_t h^{\pm}(z)=\frac{1}{2\pi i}
\p_t \left ( \oint_{\p {\sf D}}
\frac{\bar \zeta d\zeta}{\zeta -z}\right )
$$
can be found directly using a parametrization
of the family of contours $z=z(\sigma , t)$,
where $\sigma$ is a parameter along the contour.
For each component of the boundary we have:
$$
\p_t \left ( \oint
\frac{\bar z dz}{z -a}\right )=
\int \left (
\frac{\overline{z_t} z_{\sigma}+
\bar z z_{\sigma t}}{z-a}-
\frac{\bar z z_{\sigma}z_t}{(z-a)^2}\right )d\sigma =
$$
$$
\mbox{(integrating by parts)}\; =\,
\oint \left (
\frac{\overline{z_t} z_{\sigma}-
\overline{z_{\sigma}} z_{t}}{z-a}\right )
\frac{dz}{z_{\sigma}}
$$
The jump of the boundary values of the analytic
function defined by the
latter Cauchy integral is equal to
$\overline{z_t}- \overline{z_{\sigma}}z_t (z_{\sigma})^{-1}$.
Combining this with (\ref{onehand}), we obtain the
relation for differentials along the boundary curves,
\beq\label{LGE}
\p_t \overline{z(\sigma , t)} dz -
\p_t z(\sigma , t) d\bar z =
-\p_z p(z) dz
\eeq
valid for any parametrization of the contours.
In a simply-connected case with an oil sink at infinity
$p(z)=-2Q_\infty\log |w(z)|$, where $w(z)$ is a conformal map of the oil
domain to the exterior of the unit disk. Choosing $\sigma=-i\log w(z)$
Eq.(\ref{LGE}) becomes the celebrated
Laplacian growth equation
$$
\overline{z_t} z_\sigma -
  z_t \overline{ z_\sigma} =iQ_\infty.
$$
It was first derived
in \cite{DAN}.

\subsection{Dual systems of local coordinates
in the space of  multiply-connected domains}

Our next goal is to introduce special local coordinates
in the space of multiply-connected domains, which evolve linearly in
time.
The time evolution of the Cauchy transform suggests that
such coordinates are basically harmonic moments of the oil domain.

\begin{figure}[tb]
\epsfysize=8cm
\centerline{\epsfbox{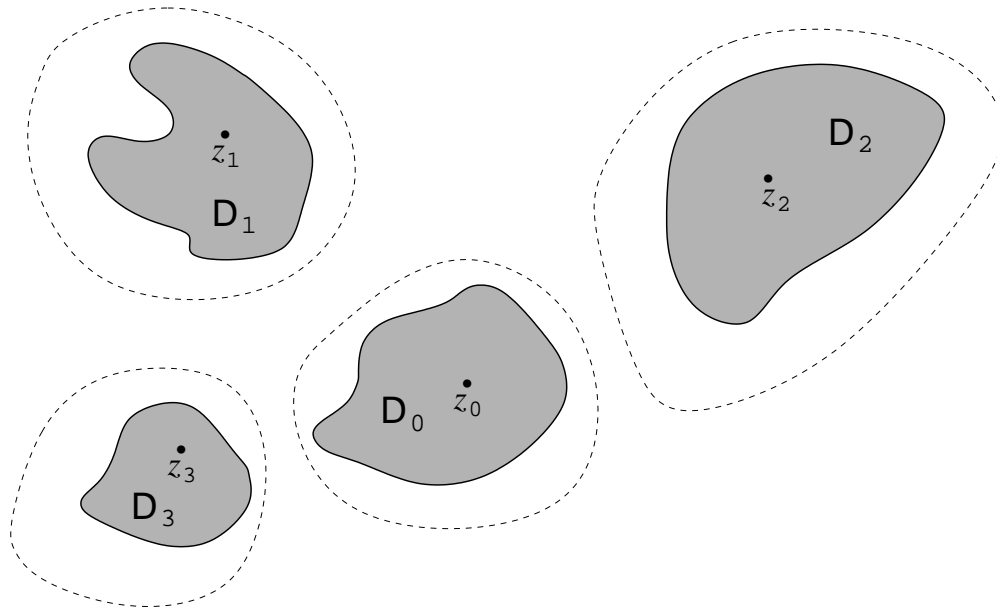}}
\caption{\sl The domain $\tilde{\sf D}_{{\sf oil}}$ is the exterior
of the regions bounded by the dashed lines.}
\label{fi:tildew}
\end{figure}

\paragraph{The proper basis of harmonic functions.} Let us consider a
time-independent domain $\tilde {\sf D}_{{\rm oil}}
\subset \Doil$ with the same connectivity as
$\Doil$, as shown in Fig.~\ref{fi:tildew},
and define  a proper basis of harmonic functions in
$\tilde {\sf D}_{{\rm oil}}$. A basis is said to be
proper if any harmonic function in $\Doil$
is representable as a linear combination
(possibly infinite) of the basis functions
such that it converges  everywhere
in $\tilde {\sf D}_{{\rm oil}}$. In the case of a single water
droplet, the basis consisting of
functions $z^{-n}$ and their conjugates is clearly a proper one.
However, this basis is no longer proper on the plane with
more than one hole.
Indeed, in this case
one has to incorporate
functions with singularities in any hole, not only
in ${\sf D}_0$, otherwise the series
converges only in some simply-connected neighborhood of infinity.

To construct a proper basis, we need
some auxiliary data.
Inside each water droplet, let us fix a point,
$z_{\alpha}\in {\Dalpha}$, which does not move with time
(Fig.~\ref{fi:tildew}).
These points may be thought of as locations of
point-like water pumps.
Without any loss of
generality, we set $z_0 =0$.

Consider the following single-valued
analytic functions in $\Doil$:
\beq\label{basis1}
\psi_k (z)=\prod_{\gamma =0}^{k-1}\frac{1}{z-z_{\gamma}}\,,
\;\;\;\;\; k\geq 0,
\eeq
where the running index $\gamma$ is understood
modulo $g+1$, i.e., it is implied that
$z_{\gamma +(g+1)m}\equiv z_{\gamma}$.
At $k=0$ we set $\psi_0 =1$. In a more
explicit form, we write
\beq\label{basis2}
\psi_{(g+1)m +\alpha}=
r^{-m}(z) \prod_{\beta =0}^{\alpha -1}(z-z_{\beta})^{-1}\,,
\eeq
where $m\geq 0$, $\alpha =0, 1, \ldots , g$
(when $\alpha =0$ the product is set to be $1$) and
\beq\label{basis3}
r(z)=\prod_{\beta =0}^{g} (z-z_{\beta})
\eeq
These functions constitute a proper basis
of single-valued analytic functions.
(As is seen from (\ref{basis2}), the domain
$\tilde {\sf D}_{{\rm oil}}$
can be choosen to be
$|r(z)|>C$ for some positive $C$.)
This basis is a simplified
version of the special Laurent-Fourier type
basis on Riemann surfaces constructed in \cite{KN}.
In the case of the single hole ${\sf D}_0$
with $z_0 =0$ it coincides with the standard one:
$\psi_k =z^{-k}$.

Any harmonic function is the real part of an analytic function.
However, in the multiply-connected case these analytic functions
are not necessarily single-valued (only their real parts
have to be single-valued).
The full basis of harmonic functions is, therefore:
$\psi_k(z)$, $\overline{\psi_k(z)}$, $k\geq 1$,
$\psi_0 =1$, and $g$ functions of the form
\beq\label{basisnu}
\ell_{\alpha}(z)=\log \left | 1-\frac{z_{\alpha}}{z}\right |^2\,,
\;\;\;\;\;
\alpha =1, \ldots , g
\eeq
which have multivalued analytic part.

An important feature of the basis $\psi_k$ is the existence
of the ``dual'' basis of differentials, $d\chi_k (z)$,
such that
\beq\label{basis4}
\frac{1}{2\pi i}\oint_{\p {\sf D}} \psi_k d\chi_n =\delta_{kn}
\eeq
Explicitly, they are given by
\beq\label{basis5}
d\chi_k (z)=
\frac{dz}{(z-z_{k-1})\psi_k (z)} =dz \prod_{\gamma =0}^{k-2}
(z-z_{\gamma})
\eeq
where $z_{k}$ is again understood as $z_{\alpha}$ where
$k=(g+1)m +\alpha$ with non-negative integer $m$
and $0\leq \alpha \leq g$. It is clear that
$\chi'_k(z)=d\chi_k / dz$, for $k\geq 1$,
are polynomials of degree $k-1$.

One can easily check that the following expansion
of the Cauchy kernel holds true:
\beq\label{basis6}
\frac{1}{\zeta -z} =
\sum_{k=1}^{\infty} \psi_k (\zeta )\chi'_k (z)
\eeq
and the series converges if $|r(z)|<|r(\zeta )|$.
In particular, the series certainly converges
if $\zeta$ is in $\Doil$ while $z$ is close enough to
any one of the points $z_{\alpha}$.
A similar expansion holds for the  logarithmic kernel:
\beq\label{basis7}
\log (\zeta -z)=\log (\zeta -z_{\alpha})-
\sum_{k=1}^{\infty}\psi_k (\zeta )\int_{z_{\alpha}}^{z}
d\chi_k
\;\;\;\;\;
\mbox{for $\zeta$ in $\Doil$ and $z$ in $\Dalpha$}
\eeq
with the same condition of convergence.

\paragraph{The harmonic moments.}

Using the expansion of the Cauchy kernel (\ref{basis6}),
let us expand the function $h^{+}(z)$ in a series,
which converges at least for $z$ close enough
to any one of the points $z_{\alpha}$. We obtain:
$$
h^{+}(z)=\sum_{k\geq 1}kT_k \chi'_k (z)
$$
where
\beq\label{im2}
T_k =-\frac{1}{\pi k}\int_{\Doil }
\psi_k (\zeta ) d^2 \zeta
\eeq
are normalized {\it harmonic moments}
of the domain $\Doil$.
(In the case of the single water droplet ${\sf D}_0$
they have the familiar form
$T_k =-\frac{1}{\pi k}\int_{\Doil }
\zeta^{-k} d^2 \zeta$.)
Their time derivatives are  read
from (\ref{hdot}):
\beq\label{Tdot}
\dot T_k = \frac{1}{k}\sum_j Q_j \psi_k (a_j)
\eeq
In case of the only sink at infinity
the r.h.s. vanishes  \cite{Rich}.
The moment $T_0$ is the area of
$\Doil$. It is infinite, but
as  long  as we need only its change,
we may equivalently consider the
  area of the complimentary domain
$$
T_0 =\frac{1}{\pi} \sum_{\alpha =0}^{g}
  \mbox{Area $(\Dalpha )$}
$$
Evidently, $\dot T_0$ is equal to
the total power of oil pumps:
$\dot T_0 =Q_{\infty} +\sum_j Q_j$.

To characterize the geometry of a multiply-connected
domain with $g+1$ boundary components one needs
$g$ extra parameters.
There are different
possibilities to choose them.
One of possible choices is as follows.
Using the expansion of the logarithmic kernel
(\ref{basis7}), or integrating the expansion
of the $h(z)$, we can represent the Coulomb
potential function in the form
\beq\label{phi2}
\phi (z)=\Phi_{\alpha} + 2{\rm Re}
\left ( \sum_{k\geq 1}kT_k \int_{z_{\alpha}}^{z}
d\chi_k \right )
\;\;\;\;
\mbox{for $z$ in $\Dalpha$}
\eeq
where $\Phi_{\alpha}=\phi (z_{\alpha})$ are integration constants.
Since $\phi (0)=0$, we set $\Phi_0 =0$.
The other integration constants, $\Phi_1 , \ldots , \Phi_g$,
may serve as the additional independent
parameters.
Clearly,
$$
\Phi_{\alpha} =\frac{1}{\pi}
\int_{\Doil} \log \left |
1-\frac{z_{\alpha}}{\zeta}\right |^2 d^2 \zeta
$$
are harmonic moments of the $\Doil$ with respect to the
functions $\ell_{\alpha}(z)$ (\ref{basisnu}).
These moments are always real.
The time derivative $\dot \Phi_{\alpha}=
\dot \phi (z_{\alpha})$ is given by
(\ref{phidot}):
\beq\label{pidot}
\dot \Phi_{\alpha}=p_{\alpha}-\sum_j Q_j
\log \left | 1-\frac{z_{\alpha}}{a_j}\right |^2
=p_{\alpha} -\sum_j Q_j \ell_{\alpha}(a_j)
\eeq

There is an alternative choice of the $g$ extra
parameters which is ``dual'' to the choice above.
Let
\beq\label{im4}
S_{\alpha}= \frac{\mbox{Area}(\Dalpha )}{\pi}
\eeq
be areas of the water droplets (divided by $\pi$), then
\beq\label{im5}
\dot S_{\alpha} =q_{\alpha}
\eeq
by the definition of $q_{\alpha}$.
Clearly, $\sum_{\alpha =0}^{g} S_{\alpha}=T_0$,
so $S_1 , \ldots , S_g$ can be taken as independent
parameters.

\paragraph{Local coordinates in the space of multiply-connected  domains.}

The basic fact from the theory of deformations of
planar domains (which we adopt without
proof in this paper) is that the parameters
$T_k , \Phi_{\alpha}$ or $T_k , S_{\alpha}$
can serve as local coordinates in the space
of planar multiply-connected domains.
This means that any deformation which preserves
these parameters is trivial and any vector field
in the space of these parameters generates a
well-defined deformation of the initial domain
(see \cite{krimaza} for details).
Recovery of the domain having these parameters constitutes the
classical inverse potential problem \cite{P.S.Novikoff}
for the multiply-connected case.
The fact that the coordinates introduced above are good ones
amounts to
local existence and uniqueness of such a domain.
An effective reconstruction, however,
is not generally
feasible and even specific examples are challenging
and merit separate attention.

The formulas for time derivatives of $T_k$, $\Phi_{\alpha}$
and $S_{\alpha}$ tell us that
any ILG flow can be represented, in these coordinates,
as a vector field with constant coefficients.

To summarize, we have introduced two systems of local coordinates
in the space of multiply-connected domains:
\begin{itemize}
\item
$T\Phi$-coordinates: the harmonic moments $T_0 , T_1 , T_2 ,
\ldots \;$ (and their complex conjugates) and
$\Phi_1 , \Phi_2 , \ldots , \Phi_g$;
\item
$TS$-coordinates: the harmonic moments $T_0 , T_1 , T_2 ,
\ldots \;$ (and their complex conjugates) and
$S_1 , S_2 , \ldots , S_g$.
\end{itemize}
Now, from
(\ref{Tdot}), (\ref{pidot}) and (\ref{im5}) it is
clear that
Richardson's result can be reformulated by saying that
the ILG dynamics with fixed pressure differences
(the $p$-representation)
is linearized in the $T\Phi$-coordinates while the ILG
dynamics with fixed pumping rates (the $q$-representation)
becomes linear in $TS$-coordinates.

\subsection{Elementary growth processes}

A linear time dependence of the local
coordinates suggests to treat any general ILG
process
as a superposition of certain ``elementary'' processes.
It is natural to associate with each elementary flow
its own time variable.

In the $p$-representation, the elementary processes are:
\begin{itemize}
\item [(p1)]
Oil is sucked from a point $a$ with the unit rate ($Q=1$),
with maintaining equal pressures in all droplets
($p_{\alpha}=0$):
$$
\dot T_0 =1, \;\;
\dot T_k =\frac{1}{k}\psi_k (a), \;\;
\dot \Phi_{\alpha}=
-\ell_{\alpha}(a)
$$
With this process we associate the time variable
$T^{(a)}$ such that
\beq\label{p1}
\frac{\p}{\p T^{(a)}}=
\frac{\p}{\p T_0}+
\sum_{k\geq 1}\frac{1}{k}
\left ( \psi_k (a) \frac{\p}{\p T_k}+
\overline{\psi_k (a)} \frac{\p}{\p \bar T_k}\right )
-\sum_{\alpha}\ell_{\alpha}(a)\frac{\p}{\p \Phi_{\alpha}}
\eeq
(the values of the coefficients follow from
(\ref{Tdot}), (\ref{pidot})).
Note that $T^{(a)}$ is the amount of oil
sucked out from the point $a$
during the process.
\item [(p2)]
Water is redistributed between the droplets by applying
the unit pressure difference between the $\alpha$-th and
the $0$-th droplets: $p_{\beta}=\delta_{\alpha \beta}$,
with no pumps in oil:
$$
\dot T_0 =\dot T_k =0, \;\;
\dot \Phi_{\beta}=\delta_{\alpha \beta}
$$
With this process we associate the time variable
$T^{(\alpha )}$ such that
\beq\label{p2}
\frac{\p}{\p T^{(\alpha )}}=
\frac{\p}{\p \Phi_{\alpha}}
\eeq
It is the amount of water injected into the 0-th
water droplet during the process.
\end{itemize}
In the $q$-representation, the elementary processes are:
\begin{itemize}
\item [(q1)]
Oil is sucked from a point $a$ with the unit rate ($Q=1$),
with water being added to the $0$-th droplet only ($q_0 =1$):
$$
\dot T_0 =1, \;\;
\dot T_k =\frac{1}{k}\psi_k (a), \;\;
\dot S_{\beta}=0
$$
In this case the vector field $\p / \p T^{(a)}$ is represented
as
\beq\label{q1}
\frac{\p}{\p T^{(a)}}=
\frac{\p}{\p T_0}+
\sum_{k\geq 1}\frac{1}{k}
\left ( \psi_k (a) \frac{\p}{\p T_k}+
\overline{\psi_k (a)} \frac{\p}{\p \bar T_k}\right )
\eeq
\item [(q2)]
Water is sucked from the $\alpha$-th droplet
and injected into the $0$-th one with the unit rate:
$q_{\beta}=-\delta_{\alpha \beta}$,
with no pumps in oil:
$$
\dot T_0 =\dot T_k =0, \;\;
\dot S_{\beta}=-\delta_{\alpha \beta}
$$
\beq\label{q2}
\frac{\p}{\p T^{(\alpha )}}=
-\frac{\p}{\p S_{\alpha}}
\eeq
\end{itemize}

In (\ref{p1}), (\ref{q1}) $\p / \p T^{(a)}$ is to be understood
not as a partial derivative but as a vector field in the space
of $g+1$ contours. By construction, it is an invariant vector
field, i.e., it does not depend on the particular basis
of harmonic functions and corresponding local coordinates.
The same is true for $\p / \p T^{(\alpha )}$.
For a general process we have:
$$
\frac{\p}{\p t}= Q_{\infty}\frac{\p}{\p T_0}+
\sum_j Q_j \frac{\p}{\p T^{(a_j)}} +
\sum_{\alpha}p_{\alpha}\frac{\p}{\p \Phi_{\alpha}}
\;\;\;\;\;
\mbox{($p$-representation)}
$$
$$
\frac{\p}{\p t}= Q_{\infty}\frac{\p}{\p T_0}+
\sum_j Q_j \frac{\p}{\p T^{(a_j)}} +
\sum_{\alpha}q_{\alpha}\frac{\p}{\p S_{\alpha}}
\;\;\;\;\;
\mbox{($q$-representation)}
$$
The vector fields and relations between
them are to be understood as acting
on any physical quantity depending on the shape of the
growing domain.
It is important to stress that
the linear superposition works only
for processes of the same type (i.e.,
either of the $p$-type or $q$-type).

At fixed positions of oil sinks ILG spans a finite-dimensional subspace
of an infinite dimensional variety of $g+1$- domains. In this subspace
$\p /\p T^{(a )}$ and $\p /\p T^{(\alpha )}$  act as partial
derivatives. For example, fix $N$ points $a_j$ and consider the variety
of contours which can be obtained from some
initial configuration of droplets
as a result of an ILG process
with oil pumps at the points $a_j$.
The resulting shape of the droplets is uniquely determined
(if no singularity occurs) by total amounts of oil
sucked out from each point.
This configuration space is $N$-dimensional, and $T^{(a_j)}$ are
local coordinates in it. Similarly, one may consider
a more general configuration space, where additional
parameters are amounts of water injected into each droplet.

\section{Analytic and algebro-geometric objects
associated to the ILG}

In this section we describe analytic and algebro-geometric
objects \cite{SS} which emerge in a description  of an  evolution of
multiply-connected domains.

\subsection{Green function, harmonic measures and period matrix}

Pressure in the oil domain is expressed  in terms of the following
objects:
\begin{itemize}
\item
$G(z,z')$:
{\it The Green function of the Dirichlet boundary problem
in $\Doil $.}
The function $G(z,z')$
is symmetric and harmonic everywhere in
$\Doil $  in
both arguments except $z=z'$ where
$G(z,z')=\log |z-z'| +\ldots $ as $z\to z'$; besides,
$G(z,z')=0$ if any
of the variables $z$, $z'$ belongs to the boundary.
The Green function obeys the equation $\Delta G(z,z') =
2\pi \delta^{(2)}(z-z')$.
\item
$\omega_{\alpha}(z)$: {\it The harmonic measure of the
$\alpha$-th boundary component.} The function
$\omega_{\alpha}(z)$
is the harmonic function in $\Doil$
such that it is equal to 1 on $\p \Dalpha$
and vanishes on the other boundary curves.
Thus the harmonic measure is the solution
to the particular Dirichlet problem.
The solution is given by
\beq
\label{periodG}
\omega_{\alpha}(z)=
-\, \frac{1}{2\pi}\oint_{\p \Dalpha}
\p_n G(z, \zeta )ds,
\;\;\;\;\;\; \alpha=0, \, 1,\ldots , g
\eeq
so the harmonic measure is the period of the
differential $\p_z G dz$.
Obviously,
the sum of the harmonic measures of all boundary
components, $\p \Dalpha$, which we call {\it cycles},
 is equal to $1$.  In what follows we consider the
linear independent functions
$\omega_{\alpha} (z)$ with $\alpha =1, \ldots , g$.
\item
$\Omega_{\alpha \beta}$: {\it The period matrix.}
Taking integrals of $\omega_{\alpha}(z)$ over
nontrivial cycles,
we define
\beq\label{periodG2}
\Omega_{\alpha \beta} =- \frac{1}{2\pi}
\oint_{\p \Dbeta }\p_n \omega_{\alpha} (\zeta )ds ,
\;\;\;\;\;  \alpha , \beta =1,\ldots , g
\eeq
The matrix $\Omega_{\alpha \beta}$ is known to be symmetric,
non-degenerate and positively-definite.
It is called the period matrix because of its
direct relation to periods of holomorphic differentials
on the Schottky double of the domain $\Doil$ (see below).
\end{itemize}

We also need the following ``modified'' objects,
which are dual, with respect to the
choice of the basis of canonical cycles
(see Sec.\,3.2) to the ones introduced above.

\begin{itemize}
\item
$\tilde G(z,z')$:
The modified Green function \cite{Gakhov} defined by
\beq\label{modgreen}
\tilde G(z,z')=G(z,z')-\sum_{\alpha , \beta =1}^{g}
\omega_{\alpha}(z)(\Omega^{-1})_{\alpha \beta}
\omega_{\beta}(z')
\eeq
This function obeys the same equation $\Delta \tilde
G(z,z') = 2\pi \delta^{(2)}(z-z')$ and integrals of
$\p_n \tilde G$ over all the cycles
$\p \D1 , \ldots , \p \Dg$ are zero.
However,
instead of being zero on the boundaries, $\tilde G$
takes there different constant values.
\item
$\tilde \omega_{\alpha}(z)$: The modified harmonic measure
is defined by
\beq\label{modharm}
\tilde \omega_{\alpha}(z)
=-2\sum_{\beta =1}^{g}
(\Omega^{-1})_{\alpha \beta }\omega_{\beta}(z)
\eeq
This is simply a linear combination of
$\omega_{\alpha}$'s with domain-dependent coefficients
such that
$$
\frac{1}{2\pi}\oint_{\p \Dalpha} \p_n \tilde \omega_{\beta} ds =
2\delta_{\alpha \beta}
$$
\end{itemize}

\paragraph{The pressure field.}

Let us demonstrate how the solution
for pressure in $\Doil$ is written
in terms of the objects just introduced.
For simplicity we do this assuming only one
sink of oil with the power $Q$
located at a point $a$ ($a =\infty$ is also
possible).

The general solution for the pressure field $p=p(z)$
with $p =p_{\alpha}$ on the boundaries reads
\beq\label{p0}
p(z)=2QG(z, a) +\sum_{\alpha =1}^{g}
p_{\alpha} \omega_{\alpha}(z) =
2Q\tilde G(z, a) -\sum_{\alpha =1}^{g}
q_{\alpha} \tilde \omega_{\alpha}(z)
\eeq
It is important to note that the rates
$q_{\alpha}$ completely determine
pressures $p_{\alpha}$ in the water droplets and vice versa.
Indeed, plugging (\ref{p0}) into (\ref{qq}), we have
the relation
$$
2Q\oint_{\p \Dalpha }
\p_{n}G(a, z)ds + \sum_{\beta =1}^{g}
p_{\beta}\oint_{\p \Dalpha }\p_n \omega_{\beta}(z) ds
=-4\pi q_{\alpha}
$$
Using (\ref{periodG}) and (\ref{periodG2}),
we can write it either as a system
of linear equations for $p_{\alpha}$
(in the $q$-representation),
\beq\label{p22}
\frac{1}{2}
\sum_{\beta =1}^{g}\Omega_{\alpha \beta}\, p_{\beta}=
q_{\alpha}- Q\omega_{\alpha}(a)
\eeq
or a system of linear equations for
$q_{\alpha}$ (in the $p$-representation),
\beq\label{p2dual}
2\sum_{\beta =1}^{g}
(\Omega )^{-1}_{\alpha \beta}\, q_{\beta}=
p_{\alpha}- Q \tilde \omega_{\alpha}(a)
\eeq
Since the $g\times g$ matrix $\Omega_{\alpha \beta}$
is non-degenerate, the system has a unique solution
which is read from the equivalent ``dual'' system.
Whichever the physical conditions
in the water droplets are,
pressure is given by
\beq\label{p4}
p(z)=
2QG(a,z) - 2Q\sum_{\alpha , \beta =1}^{g}
  \omega_{\alpha}(a)
(\Omega^{-1})_{\alpha \beta}
  \omega_{\beta}(z)   +
2\sum_{\alpha , \beta =1}^{g}
q_{\alpha}(\Omega^{-1})_{\alpha \beta}\omega_{\beta}(z)
\eeq
Specifying this formula for the elementary processes,
we have:
\beq\label{p5}
p(z)=\left \{
\begin{array}{ll}
2G(a,z) & \mbox{for (p1)}
\\
\omega_{\alpha}(z) & \mbox{for (p2)}
\end{array}
\right. ,
\;\;\;\;
p(z)=\left \{
\begin{array}{ll}
2\tilde G(a,z) & \mbox{for (q1)}
\\
\tilde \omega_{\alpha}(z) & \mbox{for (q2)}
\end{array}
\right.
\eeq

\begin{figure}[tb]
\epsfysize=5cm
\centerline{\epsfbox{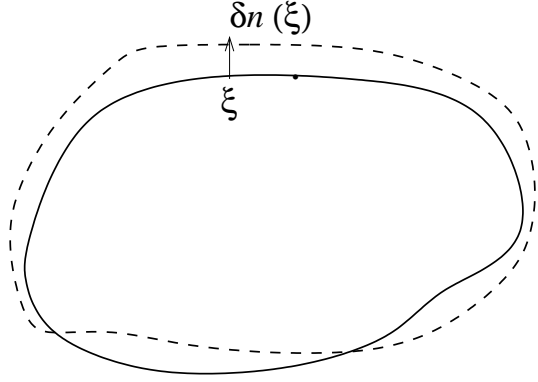}}
\caption{\sl The normal displacement of the boundary.}
\label{fi:var}
\end{figure}

\paragraph{Variational formulas.}
Variations of the Green function and
harmonic measures under infinitesimal
deformations of the domain are described by remarkable
formulas going back to Hadamard \cite{H,SS}. Let $\delta n(\xi)$
be the normal displacement (with sign) of the boundary under the
deformation counted along the normal vector at the boundary
point $\xi$, with the normal vector
looking inside $\Doil$, see Fig.~\ref{fi:var}.
The variational formulas are:
\beq\label{varG}
\delta G(z,z')=
\frac{1}{2\pi}\oint_{\p \Doil}
\p_n G(z, \xi)\, \p_n G(z', \xi)\,
\delta n(\xi) \, ds
\eeq
\beq\label{varomega}
\delta \omega_{\alpha}(z)=
\frac{1}{2\pi}\oint_{\p \Doil}
\p_n G(z, \xi)\, \p_n \omega_{\alpha}(\xi)\,
\delta n(\xi) \, ds
\eeq
\beq\label{varperiod}
\delta \Omega_{\alpha \beta}=
\frac{1}{2\pi }\oint_{\p \Doil}
\p_n \omega_{\alpha}(\xi)\, \p_n \omega_{\beta}(\xi)
\, \delta n(\xi) \, ds
\eeq
Small variations of the modified objects
(with tilde) are described
by exactly the same formulas (\ref{varG}), (\ref{varomega}),
where one should
put tilde everywhere.

These variational formulas are easy to understand.
Let us explain them on the example of the Green function.
Since the Green function $G(z,\xi)$ vanishes
if $\xi$ belongs to the old boundary,
the variation $\delta G(z,\xi)$ is equal to
the value of the new Green function
on the old boundary, i.e.,
$\delta G(z,\xi) = -\delta n(\xi)\p_n G(z,\xi)$, in the
leading order.
Now notice that $\delta G(z, \xi)$ is a {\it harmonic
function} (the logarithmic singularity cancels since it is
the same for both old and new functions)
with the boundary value
$-\delta n(\xi)\p_n G(z,\xi)$.
This function can be restored from its boundary value
by solving the Dirichlet boundary problem.
The r.h.s. of eq. (\ref{varG}) gives the result.

When the domain evolves with time, the Green function
and harmonic measures
become time dependent. The variational formulas allow one
to find time derivatives of these functions given
a local law of motion of the boundary. This is the
way how we derive partial differential equations for the
ILG below.

\subsection{The Schottky double}

The growing domain $\Doil$ is a Riemann surface
with a border.
 From mathematical point of view,
it is more convenient to work with compact
Riemann surfaces without border than
with bordered domains. Given a planar
domain with holes, like $\Doil$,
endowed with the holomorphic coordinate $z$,
it may be thought of as
a ``half'' of a closed Riemann surface.
Another half, an antiholomorphic ``copy'' of $\Doil$
with coordinate $\bar z$,
is glued to the first copy along the boundaries
$\p \Dalpha$. Besides, each copy of $\Doil$ should be
compactified by adding a point at infinity.
The resulting compact Riemann surface without
boundary is called
{\it the Schottky double}, or simply the double
of the planar bordered domain (see, e.g. \cite{SS}).

\begin{figure}[tb]
\epsfysize=7cm
\centerline{\epsfbox{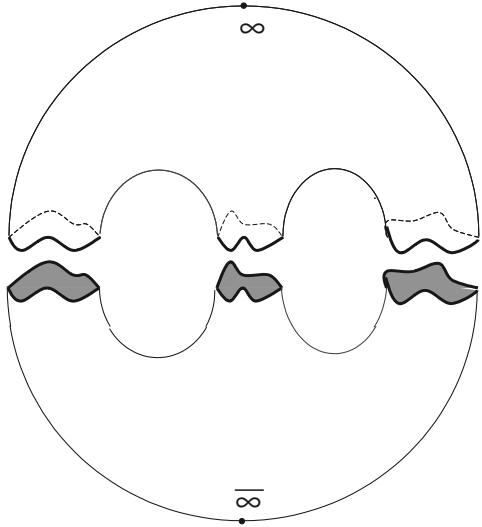}}
\caption{\sl The two halfs of the Schottky double ($g=2$).}
\label{fi:double}
\end{figure}

More precisely, the Schottly double
of a bordered surface,
is a compact Riemann surface
without boundary
endowed with an antiholomorpic involution such that the boundary
of the initial domain is the set of fixed points
of the involution.
The Schottky double of the
$\Doil$ can be
thought of as two copies
of it (``upper" and ``lower" sheets
of the double) glued along
the boundaries
$\cup_{\alpha=0}^g \p \Dalpha$, with
points at infinity added
($\infty$ and $\bar \infty$).
The holomorphic coordinate on
the upper sheet is $z$, the same as in $\Doil$,
while the holomorphic
coordinate on the lower sheet is $\bar z$.
The genus of the double is equal to the number
of water droplets minus 1.
The notion of the double was implicitly used by
Richardson in his studies of viscous flows in
multiply-connected fluid regions \cite{Rich1,Rich2}.

A meromorphic function on the double is a pair
of meromorphic functions $f,\tilde f$ in $\Doil$
such that $f(z)=\tilde f (\bar z)$ on the boundary.
Similarly, a meromorphic differential on the double
is a pair of meromorphic differentials
$f(z)dz$ and $\tilde f (\bar z)d\bar z$ such that
$f(z)dz =\tilde f(\bar z)d\bar z$ along the boundary
curves. The Schwarz reflection
principle says that any meromorphic
differential $d \nu (z)$ on the upper sheet such that
it is purely imaginary along the boundary, can be
meromorphically extended to the lower sheet as
$-d\overline{\nu (z)}$, so for each pole of such a
globally defined differential there is a ``mirror'' pole
on the opposite sheet.

To proceed, one has to choose a basis
of ${\sf a}$- and ${\sf b}$-cycles on the double
having the canonical
intersection form
${\sf a}_{\alpha} \circ {\sf a}_{\beta} =
{\sf b}_{\alpha} \circ {\sf b}_{\beta} = 0$,
${\sf a}_{\alpha} \circ {\sf b}_{\beta} = \delta_{\alpha \beta}$.
In general, for an abstract Riemann surface,
there is no preferred choice of the basis.
However, when the surface is the double of a planar
domain, like in our case, we may fix
the following two distinguished (``dual'') bases.

\begin{itemize}
\item
The ${\sf b}$-cycles are just boundaries of the
holes $\balpha =-\p \Dalpha$ for $\alpha =1, \ldots , g$.
(Note, however, the negative
{\it clockwise} orientation.)
The $\aalpha$-cycle connects the boundary of the $\alpha$-th
water droplet
with the 0-th one. To be more precise,
fix points $\xi_{\alpha}$ on the boundaries, then
the $\aalpha$-cycle starts from $\xi_{0}$,
goes to $\xi_{\alpha}$ on the ``upper''
sheet and returns following the same way on the
``lower'' sheet.
\item
In the ``dual'' basis, we interchange
${\sf a}$- and ${\sf b}$-cycles:
$\tilde \aalpha =-\balpha$,
$\tilde \balpha =\aalpha$.
The minus sign is necessary to preserve the
anti-symmetric intersection form.
\end{itemize}

These two choices
of the basic cycles correspond to
the LG dynamics in the $p$ and $q$-representations.

\begin{figure}[tb]
\epsfysize=7cm
\centerline{\epsfbox{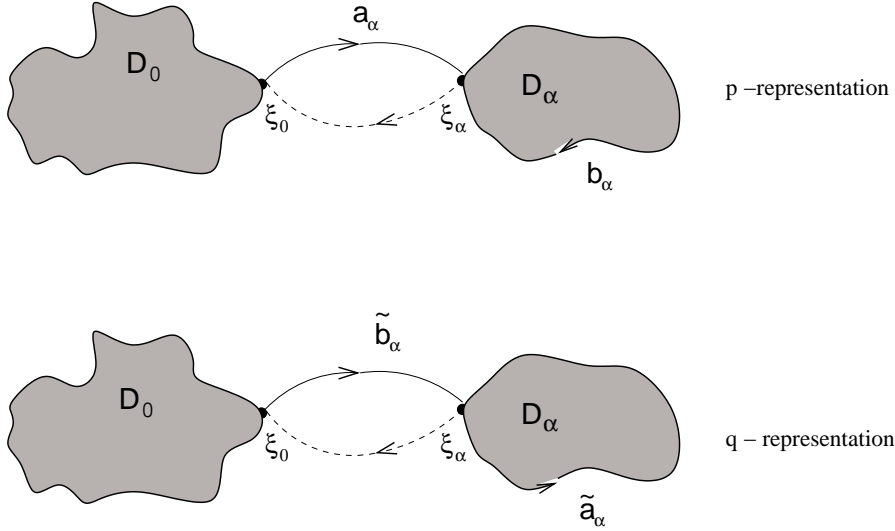}}
\caption{\sl Canonical cycles on the Schottky double
in $p$ and $q$-representations.}
\label{fi:cycles}
\end{figure}

\subsection{Differentials on the Schottky double}

\paragraph{Holomorphic differentials.}

Having fixed the basis of cycles, we can introduce
normalized holomorphic differentials
(abelian differentials of the first kind)
as differentials of holomorphic parts of the harmonic measures.
Let us represent $\omega_{\alpha}$
as the real part of a holomorphic function:
$\omega_{\alpha}=W_{\alpha}(z)+
\overline{W_{\alpha}(z)},$
where $W_{\alpha}(z)$
are holomorphic (in general multivalued)
functions in $\Doil$. The differentials
$dW_{\alpha}=\p_z \omega_{\alpha} dz$
are then holomorphic in $\Doil$ and purely imaginary
along all boundary contours. So they
can be extended holomorphically to the lower sheet
as $-d\overline{W_{\alpha}(z)}$.
They form the canonically normalized basis
of holomorphic differentials on the double
w.r.t. the ${\sf a}$-cycles:
$$
\oint_{{\sf a}_{\alpha}}\! dW_{\beta} =
\int_{\xi_0}^{\xi_{\alpha}}
dW_{\beta}(z)+
\int_{\xi_{\alpha}}^{\xi_0}
\left ( - d\overline{W_{\beta}(z)}\right )=
2 {\rm Re} \int_{\xi_0}^{\xi_{\alpha}}
dW_{\beta}(z)=
\int_{\xi_{0}}^{\xi_{\alpha}} d\omega_{\beta}
=\delta_{\alpha \beta}
$$
The matrix of ${\sf b}$-periods of these differentials
reads (cf.\,(\ref{periodG2}))
\beq\label{Talbe}
T_{\alpha \beta}=\oint_{{\sf b}_{\alpha}}
dW_{\beta} =
- \frac{i}{2}\oint_{{\sf b}_{\alpha}}\p_n \omega_{\beta} ds
=i\pi \Omega_{\alpha \beta}
\eeq

Similarly, the differentials
$d\tilde W_{\alpha}=\p_z \tilde \omega_{\alpha}
dz$,
constructed in the same way from the modified harmonic
measures, are normalized w.r.t. the
$\tilde {\sf a}$-cycles\footnote{Normalization of the holomorphic
differentials differs  by $2\pi i$ . One might work using the unified
normalization in both $p$ and $q$-representations but in that case,
as a price for the unified notation,
some artificial imaginary units enter formulas
for physical quantities.}:
$$
\oint_{\tilde {\sf a}_{\alpha}}d\tilde W_{\beta} =
2 \pi i \, \delta_{\alpha \beta}
$$
The $\tilde {\sf b}$-periods of the differentials
$d\tilde W_{\alpha}$ are:
$$
\oint_{\tilde {\sf b}_{\alpha}}d\tilde W_{\beta} =
-2(\Omega^{-1})_{\alpha \beta}
$$

\paragraph{Meromorphic differentials \cite{SS}.}

The simplest meromorphic differential on a Riemann surface
is the differential having only two simple poles
with residues $\pm 1$ (the abelian differential
of the third kind).
We will especially need the differentials whose
only simple poles are at the mirror points of the double.
They are naturally associated with the oil pumps.

Specifically, consider the differential
$dW^{(a, \bar a)}(z)=
2\p_z G(a,z)dz$ on the upper sheet.
 From the properties of the Green function it follows that
this differential has the only simple pole with residue $+1$
at the point $a$. Since along all the boundaries
$$
0=d_z G(a,z) =\p_z G(a,z)dz +\p_{\bar z}G(a,z)d\bar z =
2{\rm Re}(dW^{(a, \bar a)}(z))=0
$$
this differential can be meromorphically
extended to the lower sheet
as $-2\p_{\bar z}G(a,z)d\bar z$, and so it has
a simple pole with residue $-1$ at the mirror
point $\bar a$ on the lower sheet.
In particular, the differential
$dW^{(\infty , \bar \infty )}(z)=2\p_z G(\infty ,z)dz$
has simple poles at the two infinite points of the
Schottky double.
The differentials
$d\tilde W^{(a, \bar a)}$
are constructed in the same way
out of the modified Green function.
Note that
the so defined differential
$dW^{(a, \bar a)}$ are canonically normalized:
$$
\oint_{{\sf a}_{\alpha}}
dW^{(a, \bar a)} =0
\;\;\;\;
\mbox{that means}\;\;
{\rm Re}\int_{\xi_0}^{\xi_{\alpha}}
\p_z G(a,z)dz =0 \;\;\;\; \mbox{($p$-representation)}
$$
$$
\oint_{\tilde {\sf a}_{\alpha}}
d\tilde W^{(a, \bar a)} =0
\;\;\;\;
\mbox{that means}\;\;
\phantom{Re}\oint_{\p \Dalpha}
\p_z G(a,z)dz =0 \;\;\;\; \mbox{($q$-representation)}
$$

The abelian differentials of the second kind,
$dW^{(b)}_{k}$,
have only one pole of order $k+1$ at the point $b$.
These differentials can be explicitly defined by
expansion of the Green function or the differential
$dW^{(a, \bar a)}$ in a Taylor series in $a-b$
in the vicinity of the point $b$.
This differential is normalized, i.e.,
$\oint_{{\sf a}_{\alpha}}
dW^{(a)}_{k}=0$ but to define it uniquely one has to
fix the principal part at the pole.

Let us introduce the differentials
$dW^{(\infty )}_{k}$ with the pole at infinity.
Expanding the Green function around $\infty$
on the upper sheet, and using the basis of analytic
functions $\psi_k$ (\ref{basis1}), we write
\beq\label{expnd}
2\p_z G(a,z)dz =2\p_z G(\infty , z)dz -
\sum_{k\geq 1} \frac{1}{k}\left (
\psi_k (a) dW_{k}^{(\infty )}(z)+
\overline{\psi_k (a)}
dW_{k}^{(\bar \infty )}(z)\right )
\eeq
Here $dW^{(\infty )}_{k}$ is the normalized
differential holomorphic everywhere except infinity
(on the upper sheet) with the principal part
$$
dW_{k}^{(\infty )}(z)=k d\chi_k (z) +O(z^{-2})
\;\;\;\;
\mbox{as $z \to \infty$}
$$
On the lower sheet, this differential is regular.
Similarly,
$dW^{(\bar \infty)}_{k}$ has a pole at infinity
on the lower sheet of the double and is regular
on the upper sheet.
The differentials are defined by the following
contour integrals:
\beq\label{2kind}
dW^{(\infty )}_{k}=
\left \{
\begin{array}{c}
\displaystyle{ \frac{kdz}{\pi i}\, \p_z
\oint_{\infty} \chi_k (\zeta ) \p_{\zeta}G(z, \zeta )d\zeta}
\;\;\;\mbox{(on the upper sheet)}
\\ \\
\displaystyle{-\, \frac{kd\bar z}{\pi i}\, \p_{\bar z}
\oint_{\infty} \chi_k (\zeta ) \p_{\zeta}G(z, \zeta )d\zeta}
\;\;\;\mbox{(on the lower sheet)}
\end{array}
\right.
\eeq
The contour here encircles $\infty$ but not the point $z$.
The differential $dW^{(\bar \infty)}_{k}$ is defined as
the complex conjugated differential
$(-dW^{(\infty)}_{k})$ on the opposite sheet of the double.

The differentials introduced above
are summarized in the table:

\begin{center}

\begin{tabular}{|c|c|p{4.3cm}|c|c|}
\hline
Type & Notation & Explicit form on the upper/lower sheet &
$\displaystyle{\oint_{{\sf a}_{\beta}}}$ &
$\displaystyle{\oint_{{\sf b}_{\beta}}}$
\\
\hline
First kind & $dW_{\alpha}$ &
$\displaystyle{\phantom{\frac{A}{B}}}
\p_z \omega_{\alpha}(z) dz$ & $\delta_{\alpha \beta}$ &
$\pi i \Omega_{\alpha \beta}$\\
\cline{3-3}
& & $\displaystyle{\phantom{\frac{A}{B}}}
-\p_{\bar z} \omega_{\alpha}(z) d\bar z$ & &\\
\hline
Second kind & $dW^{(\infty )}_{k}$ &
$\phantom{\frac{\frac{\frac{A}{B}}{B} }{\frac{A}{\frac{A}{B}}}}
\mbox{see (\ref{2kind})}
$ & 0 & $
k\displaystyle{\oint_{\infty} dz \chi_k (z)
\p_z \omega_{\alpha}(z)}$
\\
\cline{3-3}
& &
$\phantom{\frac{\frac{\frac{A}{B}}{B} }{\frac{A}{\frac{A}{B}}}}
\mbox{see (\ref{2kind})}
$ & &\\
\hline
Third kind & $dW^{(a, \bar a)}$ &
$\displaystyle{\phantom{\frac{A}{B}}}
2\p_z G(a,z)dz$ & 0 & $2\pi i \omega_{\beta}(a)$ \\
\cline{3-3}
& &
$\displaystyle{\phantom{\frac{A}{B}}}
-2\p_{\bar z}G(a,z)d \bar z$
& &\\
\hline
\end{tabular}

\end{center}

\paragraph{Correspondence between   times and differentials.}

Each elementary  flow is
naturally ``coupled'' with a meromorphic or
holomorphic differential on the Schottky double.
This differential can be defined
in terms of pressure as $\p_z p(z)dz$. Equivalently it is
a unique analytic continuation of the differential
$\p_t z(\sigma , t) d\bar z -
\p_t \overline{z(\sigma , t)} dz$
to $\Doil$ from the boundary
(here $\sigma$ is any parameter on the contours).
The equivalence of the two definitions follows
from the Laplacian growth equation (\ref{LGE}) (Sec.\,2.2).
Explicitly, the coupling is:
$$
T^{(a)} \longmapsto dW^{(a, \bar a)}\,,
\;\;\;\;
T^{(\alpha )} \longmapsto dW_{\alpha}\,,
\;\;\;\;
T_k \longmapsto  dW^{(\infty )}_{k}
$$

In particular, $T^{(\infty )}=T_0 +\mbox{const}$, corresponding to
an oil sink at infinity, is coupled with the
differential $dW^{(\infty , \bar \infty )}$.
The abelian integral
\beq\label{abint}
W(z)=W^{(\infty , \bar \infty )}(z)=
\int_{\zeta_0}^{z}dW^{(\infty , \bar \infty )}
\eeq
where $\zeta_0$ is an arbitrary fixed point in $\Doil$,
has a constant real part along each component
of the boundary and
$W(z)=\log z +O(1)$ as $z \to \infty$.
For the process (q1) with the sink at infinity
the function $w(z)=e^{W(z)}$ conformally maps the
domain $\Doil$ onto the exterior of a disk
of radius $r_0 = |w(\xi_0)|$ ($\xi_0$ is any point
on ${\sf b}_0$)
with $g$ concentric arcs removed. Radii and
endpoints of the arcs depend on time.
The radius, $r_{\alpha}$, is  related to pressure
in the $\alpha$-water droplet by
$p_{\alpha}=2\log (r_{\alpha}/r_0 )$.

\section{Equivalence of the ILG and
Whitham dynamics}

We  describe an evolution of the
Riemann surface through a time dependence
of the canonical abelian differentials and
  abelian
integrals on the Schottky double.
One may realize them as holomorphic
or meromorphic functions of the coordinate $z$
on the physical plane. These functions encode the
shape of the interface at any instant of time.
In this respect they serve as substitutes for the
time-dependent conformal map to the unit disk
in case of a single droplet.

We represent the
dynamical equations of the ILG  as relations between variations of
the abelian integrals  under simultaneous action of different oil and
water pumps. Treating $T^{(a)}$ and $T^{(\alpha )}$ as an independent
``time" variables the evolution takes the form  of a hierarchy of partial
differential equations in many variables.
We recognize them as  a  universal Whitham hierarchy
\cite{kri1,kri2}
in the (extended) moduli space of
genus $g$ Riemann surfaces (see Sec. 4.3).

\subsection{Exchange relations}

Let us calculate the time derivative
of the Green function
$G(a,b)$ in the
elementary process with
the oil pump at some third point $c$.
Using the Hadamard formula (\ref{varG}),
and the fact that for this process
$\delta n(\xi )=-\frac{1}{2}\p_n G(c, \xi )\delta
T^{(c)}$,
we have:
$$
\frac{\p G(a,b)}{\p T^{(c)}}=
-\frac{1}{4\pi}\oint_{\p \Doil}
\p_n G(a, \xi)
\p_n G(b, \xi)
\p_n G(c, \xi) ds
$$
Remarkably, the result is {\it symmetric} under
all permutations of the points $a,b,c$. For a single connected domain
this equation  has been obtained in Ref. \cite{MarWZ}. In a similar way,
we find, using (\ref{varomega}), (\ref{varperiod}):
$$
\frac{\p \omega_{\alpha}(b)}{\p T^{(c)}}=
-\frac{1}{4\pi}\oint_{\p \Doil}
\p_n \omega_{\alpha}(\xi )
\p_n G(b, \xi)
\p_n G(c, \xi) ds
$$
and
$$
\frac{\p \Omega_{\alpha \beta}(b)}{\p T^{(c)}}=
-\frac{1}{4\pi}\oint_{\p \Doil}
\p_n \omega_{\alpha}(\xi )
\p_n \omega_{\beta}(\xi )
\p_n G(c, \xi) ds
$$
as well as analogous formulas for derivatives
w.r.t. $T^{(\alpha )}$. Observing the symmetry
of the right hand sides of these equations,
one may write them as local relations connecting
the time derivatives
of the Green function and harmonic measures.

In this way we obtain the following fundamental
{\it exchange relations}:
\beq\label{ex1}
\frac{\p G(a,b)}{\p T^{(c)}}=
\frac{\p G(b,c)}{\p T^{(a)}}=
\frac{\p G(c,a)}{\p T^{(b)}}
\eeq
\beq\label{ex2}
\frac{\p \omega_{\alpha}(a)}{\p T^{(b)}}=
\frac{\p \omega_{\alpha}(b)}{\p T^{(a)}}
\eeq
\beq\label{ex3}
\frac{\p \omega_{\alpha}(a)}{\p T^{(\beta )}}=
\frac{\p \omega_{\beta }(a)}{\p T^{(\alpha)}}
\eeq
which encode dynamics of the Laplacian growth
with zero surface tension.
There is also the relation which connects
derivatives of the Green function and harmonic
measure:
\beq\label{ex4}
2\frac{\p G(a,b)}{\p T^{(\alpha )}}=
\frac{\p \omega_{\alpha}(a)}{\p T^{(b)}}
\eeq
In the same way, one may extend this list
to include derivatives of the period matrix:
\beq\label{ex5}
2\frac{\p \omega_{\beta}(a)}{\p T^{(\alpha)}}=
\frac{\p \Omega_{\alpha \beta}}{\p T^{(a)}}\,,
\;\;\;\;
\frac{\p \Omega_{\alpha \beta}}{\p T^{(\gamma )}}=
\frac{\p \Omega_{\beta \gamma}}{\p T^{(\alpha)}}=
\frac{\p \Omega_{\gamma \alpha}}{\p T^{(\beta )}}
\eeq
For the dual objects, the relations
(\ref{ex1}) - (\ref{ex4}) remain the same.

\subsection{The Whitham equations}

Whitham equations are holomorphic parts of the
exchange relations. Taking, for instance, $\p_z$
of both sides of the exchange relation
$\p_{T^{(b)}}G(a,z)=\p_{T^{(a)}}G(b,z)$, we get
$\p_{T^{(b)}}dW^{(a,\bar a)}(z)=\p_{T^{(a)}}dW^{(b,\bar b)}(z)$
which is an equation of the Whitham hierarchy.
The full list of Whitham equations obtained in this
way reads:
\beq\label{W1}
\frac{\p}{\p T^{(a)}}dW^{(b, \bar b)}(z)=
\frac{\p}{\p T^{(b)}}dW^{(a, \bar a)}(z)
\eeq
\beq\label{W2}
\frac{\p}{\p T^{(\alpha )}}dW^{(a, \bar a)}(z)=
\frac{\p}{\p T^{(a)}}dW_{\alpha}(z)
\eeq
\beq\label{W3}
\frac{\p}{\p T^{(\alpha )}}dW_{\beta}(z)=
\frac{\p}{\p T^{(\beta )}}dW_{\alpha}(z)
\eeq
The derivatives are taken at constant $z$.
The list can be further enlarged by adding
the equations containing $T_k$-derivatives:
$$
\frac{\p}{\p T_n}dW^{(\infty )}_{k}(z)=
\frac{\p}{\p T_k}dW^{(\infty )}_{n}(z)
$$
and so on. They can be obtained from the generating
equations (\ref{W1}) - (\ref{W3}) by expending them
in a series around infinity.

We note that there are a few other equivalent ways to
write the Whitham equations \cite{kri2}. Let us present
a more invariant formulation, which does not rely on the
choice of the distinguished local coordinate $z$ in $\Doil$.
The Whitham equations in the invariant form
are naturally written in an extended
``moduli space'' of contours. Points of this extended
space are sets of data of the form
$$\Bigl (\mbox{$g+1$ nonintersecting boundary contours}; \,\,
\mbox{a point in $\Doil$}\Bigr )
$$
Local coordinates in this space
are parameters $T\Phi$ or $TS$ and a local coordinate
$\lambda$ in the domain $\Doil$.

The local coordinate
$\lambda$ may depend on the  shape of the domain.
In particular, one may choose $\lambda$ to be one of
the abelian integrals, say $W(z)=W^{(\infty , \bar \infty )}(z)$ with
  $\oint_{\balpha}dW=0$  (\ref{abint}).
Then
the function $w(z)=e^{W(z)}$ is single-valued in $\Doil$.
It is a good local coordinate in $\Doil$ everywhere except
for the points where $dW^{(\infty , \bar \infty )}=0$.
It can be shown that all these points
belong to the boundaries ${\sf b}_1 ,
\ldots , {\sf b}_g$ and there are exactly
two such points on each boundary. Under the conformal map
$w(z)$ these points are taken to the endpoints
of the concentric arcs (see the end of Sec.\,3.3).
Treating all other abelian integrals as functions
of $W$ rather than $z=z(W,T)$,
$$
W^{(a,\bar a)}(W,T)=\int_{\zeta_0}^{z(W,T)}
dW^{(a,\bar a)}
$$
we represent
the Whitham equations
(\ref{W1}) - (\ref{W3}) in the form
\beq\label{W101}
\frac{\p W^{(A)}}{\p T^{(B)}}-
\frac{\p W^{(B)}}{\p T^{(A)}}+
\bigl \{
W^{(A)}, \, W^{(B)} \bigr \} =0
\eeq
where $A$ stands for $a$ or $\alpha$ and
$$
\bigl \{
W^{(A)}, \, W^{(B)} \bigr \} :=
\frac{\p W^{(A)}}{\p W}
\frac{\p W^{(B)}}{\p T_0}-
\frac{\p W^{(B)}}{\p W}
\frac{\p W^{(A)}}{\p T_0}
$$
has the form of ``Poisson brackets''.
These equations are consistency conditions for
the system of evolution equations
\beq\label{W102}
\frac{\p z}{\p T^{(A)}}=\bigl \{ W^{(A)}, \, z \bigr \}
\eeq

Eqs. (\ref{W101},\ref{W102}) constitute the Whitham universal hierarchy of
the soliton theory.

\subsection{Whitham equations in soliton theory}

\paragraph{Whitham equations as modulation equations.}
Integrable partial differential and difference equations
of soliton theory are known to possess a rich family
of periodic exact solutions depending
on continuous parameters.
To be more definite, we start our discussion with
a $(1+1)$-dimensional integrable evolution equation of the form
$u_t =P(u, u_x , \ldots )$ (e.g.,  the KdV equation). Exact periodic
solutions have the form
$$
u (x,t) = u_0 ({\bf U}x +{\bf V}t + {\bf Z} \, |\, I)
$$
where ${\bf U}$,
${\bf V}$, ${\bf Z}$
are $g$-dimensional constant vectors
with components
$\{U_{\alpha}\}=(U_1 , \ldots , U_g)$, etc,
and $u_0 ({\bf Z})$ is a periodic function of
any component $Z_{\alpha}$. This function,
and  all the vectors, depend on the set
of parameters $I=(I_1, \ldots , I_M)$.
Each periodic solution can be constructed starting from a Riemann
surface.
$I$ stands for the moduli of a Riemann surface and the vectors ${\bf U}$,
${\bf V}$ are ${\sf b}$-periods of certain normalized meromorphic
differentials, $dW^{(x)}$ and $dW^{(t)}$, on the
Riemann surface, with prescribed singularities at infinity \cite{Novikov}.

In a number of
physical problems one is interested  in
slowly modulated
periodic solution, rather than just periodic.
A nonlinear WKB method or the Whitham averaging method allows one to
construct more general solutions of the same integrable equation using
the function
$u_0$ as a leading term of the asymptotic expansion
$$
u(x,t)= u_0 \Bigl (\varepsilon^{-1}{\bf S}(X,T) +
{\bf Z}(X,T)\, \bigl | \, I(X,T)\Bigr )+
\varepsilon u_1 (x,t)+\varepsilon^2 u_2 (x,t)+\ldots
$$
where $\varepsilon$ is a small parameter and
the parameters $I$ now depend on the
{\it slow variables} $X=\varepsilon x$,
$T=\varepsilon t$.
The original variables $x,t$ are called {\it fast
variables}.
If the vector-valued function
${\bf S}$ obeys the equations
$$
\p_X {\bf S}={\bf U}(I(X,T))\,,
\;\;\;\;\;
\p_T {\bf S}={\bf V}(I(X,T))
$$
then the leading term agrees with the original solution
up to first order in $\varepsilon$. All the higher corrections
can be found by solving non-homogeneous linear equations
whose homogeneous part is the original
equation linearized on the background of the
exact solution $u_0$ \cite{Whitham}.

We see that the so constructed
solution $u(x,t)$ describes the original
fast periodic oscillations, modulated, on a larger
scale, by a slow drift in the space of
exact periodic solutions. The equation, which
describes the drift  $I(X,T)$ are  called Whitham equations.
For the particular example discussed above the equation to determine
$I(X,T)$, written in a proper local parameter reads
$\p_X dW^{(t)}=\p_T dW^{(x)}$.
This form of the Whitham
equations was first observed in \cite{FFM}
for the KdV hierarchy. Implicitly, through the
dependence of the canonically normalized
differentials on the slow variables, they describe the
drift in the moduli space of Riemann surfaces and thus the
dependence $I(X,T)$.
The Hamiltonian approach to Whitham equations
for $(1+1)$-dimensional systems was developed in \cite{DN}.
A universal
  Whitham  hierarchy  in a
  general setting of multi-dimensional integrable equations was suggested
in ref. \cite{kri1}.
An invariant formulation of the Whitham hierarchy,
independent on the choice of local coordinates,
was given in \cite{kri3}.

The idea of \cite{kri1} was to obtain equations
describing the slow drift in the space of exact solutions
from the condition that next-to-leading terms of the asymptotic
series $u(x,t)$ be uniformly bounded on large scales.
In general, the asymptotic series becomes unreliable
on scales of order $\varepsilon^{-1}$, i.e., the corrections
become large. The main result of \cite{kri1} is that
the Whitham equations follow if one requires that
just the next term of the series,
$u_1(x,t)$, be uniformly bounded for all $x,t$.

\paragraph{The universal Whitham hierarchy.}
In a more general multi-dimensional
hierarchy of soliton equations (like
the KP hierarchy or the 2D Toda lattice hierarchy,
or their difference counterparts), one
has a family of ``times'' $t_A$ and a family
of ``potentials'' $u^{(N)}$, evolving with
  times $t_A$.
(Here $A$ and $N$ belong to a
case-dependent, generally infinite  set of indices.)

Exact periodic solutions of the hierarchy
are constructed from a given
time-independent Riemann surface $\Gamma$
with some additional data on it.
With each time $t_A$ one associates a meromorphic
differential, $dW_A$, on $\Gamma$
normalized with respect to, say, ${\sf a}$-cycles:
$\oint_{\aalpha}dW_A =0$.
Let ${\bf U}^{(A)}$ be the vector of
${\sf b}$-periods of this differential:
$$
U^{(A)}_{\alpha}=
\oint_{\balpha} dW_A
$$
Then the exact solution has the form
($u$ is one of $u^{(N)}$'s):
$$
u(\{t_A\})=u_0 \Bigl (
\sum_A {\bf U}^{(A)}(I) t_A + {\bf Z}(I)
\, \Bigr | \, I\Bigr ) + c_0 (I)
$$
where $u_0$ is a certain oscillating periodic function
(the second logarithmic derivative
of the Riemann $\theta$-function) and $c_0$ is a constant.
Like in the previous example, one may try to
construct a more general oscillating, but not periodic solutions
with slowly varying parameters
\beq\label{W4}
u= u_0 \Bigl (\varepsilon^{-1}{\bf S}(\{ T_A \}) +
{\bf Z}(\{ T_A \})\, \Bigr | \, I(\{ T_A \})\Bigr )+
c_0 (I( \{T_A \}))+
\varepsilon u_1 (\{t_A \})+ \ldots
\eeq
Here $T_A =\varepsilon t_A$ are slow times and
${\bf S}$ is a vector function
such that $\p_{T_A}{\bf S}={\bf U}^{(A)}(I(\{T_A \}))$.
The uniform boundness of the first correction $u_1$
for all times implies
the hierarchy of Whitham equations \cite{kri2},
\beq\label{W5}
\frac{\p W_A}{\p \lambda}\left (
\frac{\p W_B}{\p T_C}-
\frac{\p W_C}{\p T_B}\right )+
\frac{\p W_B}{\p \lambda}\left (
\frac{\p W_C}{\p T_A}-
\frac{\p W_A}{\p T_C}\right )+
\frac{\p W_C}{\p \lambda}\left (
\frac{\p W_A}{\p T_B}-
\frac{\p W_B}{\p T_A}\right )=0
\eeq
valid for all possible
values of the indices $A,B,C$. Here $\lambda$ is
any local parameter, all the abelian integrals
being regarded as functions of $\lambda$.
Choosing one of the indices, say $C$, to be
$(\infty , \bar \infty )$ and
$\lambda =W=W^{(\infty , \bar \infty )}$,
one gets the Whitham equations in the form (\ref{W101}).

While averaging the solution (\ref{W4})
over fast oscillations, $\bigl < u_0 \bigr >$ vanishes
$$
\bigl < u \bigr > (\{T_A \})= c_0 (\{T_A \})
$$
In the context of Laplacian growth,
$T_A$ is $T^{(a)}$ or $T^{(\alpha )}$, and
$c_0$
is the Green function $G(a,b)$. Here the points $(a,b)$ label the
potential. Thus the Laplacian growth can be thought
of as a physical realization of the slow drift
in the moduli space of Riemann surfaces.

\section{Special classes of solutions of the
Whitham hierarchy and Laplacian growth of
algebraic domains}

In this section we briefly discuss particularly important  families of
growing domains. They correspond to special    solutions to the
Whitham hierarchy  called ``algebraic orbits'' \cite{kri2}.

\subsection{Algebraic and abelian domains}
Let us recall that the time derivative of the Cauchy transform  of the
oil domain  (\ref{ht}) is a globally defined rational function.
Consider the class of domains whose Cauchy transform
is  a single globally
defined meromorphic (i.e., rational) function in the plane.
In other words, each function $h_{\alpha}^{+}$ defined originally in the
domain $\Dalpha$ is extendable to a single  rational function defined
everywhere in the plane, same for different $\alpha$. In this case one may
forget about the index
$\alpha$ and deal with the single function $h(z)$.
Eq.\,(\ref{ht}) tells us that if the initial
fluid region is from this class, then
it remains to be in this class in the process
of the LG evolution.
The evolution may only add new poles or change
residues of the existing ones.
As is pointed out in
\cite{Rich2}, in order to prepare such an initial
condition, one may inject oil through  points
into a cell initially filled by water.

The domains whose Cauchy transform is a globally defined
rational function are called {\it algebraic} \cite{EV} or
{\it quadrature domains} \cite{Ahar-Shap}.
Some illustrative examples
in the multiply-connected case can be found in \cite{Rich2} and
\cite{Crowdy1}.
This class appears to be
quite representative and important since any domain
with smooth boundary components can be
approximated by quadrature domains (see \cite{Gustafsson} for
the proof).
In the simply-connected
case, the quadrature domains are images
of the unit disk under conformal maps given by
rational functions. Their time evolution is
described by rational solutions of the Laplacian
growth equation. These solutions are sometimes known to develop
cusp-like singularities within finite time \cite{SB}.

A more general class of domains can be defined
by imposing the above condition not on
the $h_{\alpha}^{+}$ itself but on its $z$-derivative.
Namely, suppose that each differential
$dh_{\alpha}^{+}(z)$ is extendable
to a meromorphic differential in the plane,
and they coincide for different $\alpha$'s.
In this case $h^+$ itself may be a multivalued analytic
function with logarithmic branch points.
In \cite{EV}, such domains were called
{\it abelian domains}. They can be produced from
the quadrature domains by the oil sucking from
linearly extended sinks.
In the simply-connected case,
their evolution is described by logarithmic solutions
of the Laplacian growth equation \cite{How,M1}.

\subsection{The Schwarz function}

For domains with analytic boundaries, and
for algebraic domains in particular,
the Cauchy transform allows one to introduce
the Schwarz function of the boundary contours,
which proved to be very useful for analyzing
the LG dynamics
in the simply-connected case \cite{MWZ}.
Given a closed contour on the plane,
the {\it Schwarz function}
\cite{Ahar-Shap,Davis} is defined as the analytic
continuation of the function $\bar z$ away from the
contour. Let us denote it by $S(z)$.
According to the definition,
$S(z)$ is a function analytic in some neighborhood
of the curve such that
\beq\label{Schwarz}
S(z)=\bar z \;\;\;\;\;
\mbox{on the curve}
\eeq
 From the continuity of the Cauchy transform we have
$\bar z = h_{\alpha}^{+}(z)- h^-(z)$
on the $\alpha$-th boundary,
so $S^{( \alpha )}(z) =h_{\alpha}^{+}(z)- h^-(z)$
is the Schwarz function of the $\alpha$-th boundary curve.
In general all these functions are different.
However,
as it directly follows from the definition,
all the boundary contours of algebraic or abelian domains
have {\it a common Schwarz function},
$S(z)=S^{( \alpha )}(z)$ for any $\alpha$, and the
differential $dS(z)$ is meromorphic in $\Doil$, i.e.,
it has there only a finite number of isolated poles.

In the case of algebraic or abelian domains,
one may decompose the Schwarz function into the sum
$S(z)=S_{+}(z)+S_{-}(z)$,
where the function $S_{+}=h_{\alpha}^{+}(z)$
(for any $\alpha$) is analytic inside water
droplets while $S_{-}=-h^-(z)$
is analytic in $\Doil$ and vanishes at infinity.
Combining the time derivatives of
$h_{\alpha}^{+}$ and $h^-$
(see (\ref{hdot}) and (\ref{htildedot})),
we find the time derivative
$\dot S(z)=\p_t S(z;t)$
of the Schwarz function
at constant $z$:
\beq\label{Sdot}
\dot S(z)=\dot h_{\alpha}^{+}(z)-\dot h^-(z)
=-\p_z p(z)
\;\;\;\;
\mbox{(for $z$ in $\Doil$)}
\eeq
where by $h_{\alpha}(z)$ we mean the analytic continuation
of this function to $\Doil$.

In $\Dalpha$'s, the analytically
continued function $S(z)$ has more complicated
singularities. In case of general position, they are
branch points $\eta_i$ of order two (with cuts between them).
One may think of the algebraic curve underlying the
solution to the LG problem as the Riemann surface of
the function $S(z)$.
The Whitham equations can then be equivalently represented
in the form of equations for the endpoints of the cuts,
which describe their time dependence
$\eta_i =\eta_i (T)$.
The equations are:
\beq\label{W201}
\frac{\p \eta_i}{\p T^{(A)}}=
\left ( \frac{d W^{(A)}(z)}{dW(z)}\right )_{z=\eta_i}
\frac{\p \eta_i}{\p T_0}
\eeq
where the coefficients in front of $\p \eta_i /\p T_0$
in the r.h.s. are expressible as (in general quite complicated)
functions of the $\eta_k$'s.
Basically these
equations mean that the Schwarz function takes finite
values at the branch points $\eta_i$.

\subsection{The generating differential}

For the class of algebraic domains, one may
introduce a distinguished meromorphic differential
on the Schottky double.
Recall that the Schwarz function $S(z)$ is a meromorphic
function in $\Doil$, if $\Doil$ is an algebraic domain.
Therefore, one may treat $S(z)$ as a function on the
Schottky double extending it to the lower sheet as
$\bar z$.
In this case the differential
$d{\cal S}=S(z)dz$
is extendable to a meromorphic differential
on the double. Its explicit form on the lower sheet
is $\bar z d\overline{S(z)}$.

\begin{figure}[tb]
\epsfysize=3cm
\centerline{\epsfbox{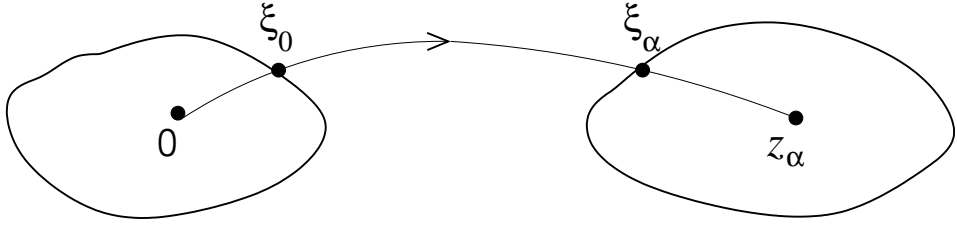}}
\caption{\sl The integration path from $z_0= 0$ to $z_{\alpha}$.}
\label{fi:path}
\end{figure}

The differential $d{\cal S}$
plays an important role
in the theory. We call it {\it the generating differential}
because it generates the complete set of local coordinates
in the space of algebraic multiply-connected domains.
Indeed, the harmonic moments are
\beq\label{tttk}
T_k =-\,\frac{1}{2\pi i k}\oint_{\p \Doil}
\psi_k (z) S(z)dz
\eeq
The areas of water droplets are periods of this differential
over ${\sf b}$-cycles:
\beq\label{f2}
S_{\alpha}=-\frac{1}{2\pi i} \oint_{\balpha}
d{\cal S}
\eeq
The ``dual'' variables, $\Phi_{\alpha}$,
are expressed through periods of the same differential
over the ${\sf a}$-cycles (or, what is the same,
over $\tilde {\sf b}$-cycles).
To show this, we write:
$$
\begin{array}{c}
\Phi_{\alpha} =
\displaystyle{
\int_{0}^{z_{\alpha}}d\phi (z)=
2\, {\rm Re}\int_{0}^{z_{\alpha}}h(z) dz =}
\\ \\
\displaystyle{
= 2\, {\rm Re} \left (
\int_{0}^{\xi_0 }h_{0}^{+}(z) dz
+\int_{\xi_{\alpha}}^{z_{\alpha}}h_{\alpha}^{+}(z) dz
+\int_{\xi_0 }^{\xi_{\alpha}}(\bar z + h^{-}(z))dz
\right )}
\end{array}
$$
where the integral from $0$ to $z_{\alpha}$
goes along a path containing the two points and
$\xi_0$, $\xi_{\alpha}$ are intersection points
of the path with the boundary curves ${\sf b}_0$,
$\balpha$ (see Fig.~\ref{fi:path}).
Adding and subtracting
$\int_{\xi_0 }^{\xi_{\alpha}}h^{+}(z) dz$
(which is well-defined for algebraic domains),
we rewrite this in the form
$$
\Phi_{\alpha}=2 {\rm Re}
\int_{0}^{z_{\alpha}}S_{+}(z)dz -\Pi_{\alpha}
$$
where
\beq\label{period3}
\Pi_{\alpha}=
2{\rm Re}\, \int_{\xi_0}^{\xi_{\alpha}}
(S(z) -\bar z) dz
=\int_{\xi_0}^{\xi_{\alpha}}
\left (S(z)dz -\bar z d\overline{S(z)}\right )
=\oint_{\aalpha} d{\cal S}
\eeq
Note that $\Pi_{\alpha}$ does not depend
on the points $\xi_0$, $\xi_{\alpha}$.
Finally, the times $T^{(a)}$ associated with
a sink of oil at the point $a$ are expressed as
\beq\label{ttta}
T^{(a)}=-\frac{1}{2\pi i}\oint_{{\sf c}_a}
d{\cal S}
\eeq
(Here ${\sf c}_a$ is a small contour encircling
the point $a$.)

The table of differentials from Sec. 3.3
can be continued by including the
generating differential:

\begin{center}

\begin{tabular}{|p{2cm}|c|p{4.3cm}|c|c|}
\hline
Type & Notation & Explicit form on the upper/lower sheet &
$\oint_{{\sf a}_{\alpha}}$ & $\oint_{{\sf b}_{\alpha}}$\\
\hline
Generating differential & $d{\cal S}$ &
$\phantom{ \frac{ \frac{A}{B} }{\frac{A}{B} }}
S(z) dz$ & $\Pi_{\alpha}$ &
$-2\pi i S_{\alpha}$
\\
\cline{3-3} & & $
\phantom{ \frac{ \frac{\frac{A}{B}}{B} }{\frac{A}{
\frac{A}{B}} }}
\bar z d\overline{S(z)}$ & &\\
\hline
\end{tabular}

\end{center}

Moreover, partial derivatives of the generating
differential w.r.t. the times
$T^{(a)}$, $T^{(\alpha )}$ coincide
with the canonical meromorphic differentials:
\beq\label{st1}
\frac{\p S(z)}{\p T^{(a)}}\, dz
=-dW^{(a, \bar a)}(z)\,,
\;\;\;\;\;
\frac{\p S(z)}{\p T^{(\alpha )}}\, dz
=- dW_{\alpha}(z)
\eeq
This follows from (\ref{Sdot}) after substituting
the pressure filed for the elementary processes
in terms of the Green function and harmonic measures
(see (\ref{p5})).
For algebraic domains, the Whitham
equations follow from the existence of the
generating differential and (\ref{st1}).

The generating differential can be represented as
\beq\label{st2}
d{\cal S}=d\Lambda -
\sum_A T^{(A)} dW^{(A)}
\eeq
where $d\Lambda$ is a
differential with
  $T^{(A)}$-independent
singularities. For algebraic orbits, it is a fixed meromorphic
differential (possibly with time-independent jumps). In more general
cases $d\Lambda$ has more complicated analytic properties.
Presumably, it can be defined as a solution to
a $\bar \p$-problem.
The expansion (\ref{st2}) and equation
$d{\cal S}(\lambda_s)=0$
for all zeros $\lambda_s$ of the differential
$dz$ on the
lower sheet of the Schottky double, where it has the form
$d\overline{S(z)}$,
are
key relations which imply (\ref{st1}). Indeed, from
the latter condition it follows that the differential
$\p_A d{\cal S}$ has no singularities at the
points $\lambda_s$.
Then (\ref{st2}) implies that this differential
has the same singularities and periods as $-dW^{(A)}$.
Hence, they do coincide.

General algebraic orbits of Whitham equations
for higher genus Riemann surfaces in the sense
of \cite{kri2} correspond to
the case when the $z$-derivative of the Schwarz function
extends to a meromorphic function on the double
(equivalently, when the differential $dS(z)$ extends
to a meromorphic differential on the double).

\section{Conclusion}

In short, the main message of this work is that
the variables in which the Laplacian growth
with zero surface tension becomes linear,
for arbitrary connectivity of the growing domain
and arbitrary configuration of pumps,
are the Whitham ``times"
defined in \cite{kri2}.
The latter are special local coordinates on the extended
moduli space of Riemann surfaces.
Conservation or linear dependence on time
of harmonic moments of the growing domain,
known before as a characteristic feature of the idealized
Laplacian growth, is a particular case of this
result.

The Whitham equations
are partial differential
equations for canonical holomorphic and meromorphic
differentials on Riemann surfaces regarded as functions
of the local coordinates in the moduli space. Solutions
to the Whitham equations allow one to find the differentials
and  abelian integrals as functions of time
and reconstruct dynamics of
  the interface.

The Whitham equations are often regarded as integrable ones,
though not in the Liouville sense.
When speaking about integrability of Whitham equations,
one means mainly a possibility to actually integrate
them by representing a solution
in the form of an implicit function of
independent variables (the hodograph method).

The  Whitham
equations appear in soliton theory in different contexts.
First, when one
looks for solutions  of soliton equations other than periodic.
At some
regimes, these solutions  are well approximated,
on small space-time scales,
by the periodic exact solutions of the
algebro-geometric type.
When fast oscillations of the periodic
solutions are averaged or
smoothed out, Whitham equations appear
as modulation equations written
for moduli of the Riemann surface parametrizing
the algebro-geometric periodic solutions.
An important special case of
Whitham equations appears if one neglects dispersion  in nonlinear
soliton equations. The latter case provides the most direct
link to the Laplacian growth
of simply-connected domains. This link was explored in Ref.\cite{MWZ}.
What is perhaps the most important conclusion,
the Whitham equations describe
a proper evolution of the Riemann surface built upon a growing interface.

The relation between the  growth  problem and
modulated periodic solutions to soliton equations is two-fold.
The Laplacian growth may serve as
a simple illustrative physical model of the Whitham dynamics of
complex curves. Vice versa, the methods developed
in soliton theory may help to understand
growth in a singular (turbulent) regime, i.e., in a vicinity of
cusp formation or coalescence and break-up of droplets, providing
an effective account of the surface tension effects near
singular points.

\section*{Acknowledgments}

We are indebted to O.Agam, E.Bet\-tel\-heim, V.Ka\-za\-kov,
A.Mar\-sha\-kov, R.Theo\-do\-res\-cu
for useful discussions and N.Am\-burg for help with figures.
The work of I.K. was supported by
NSF grant DMS-01-04621.
M.M.-W. and A.Z. were supported
by the LDRD project 20020006ER ``Unstable
Fluid/Fluid Interfaces" at Los Alamos National Laboratory.
P. W. was supported by the NSF MRSEC Program under
DMR-0213745, NSF DMR-0220198 and by Humboldt foundation.
The work of A.Z. was also supported in part by RFBR grant
03-02-17373 and by grant for support of scientific schools
NSh-1999.2003.2.
A.Z. is grateful to CNLS at Los Alamos National Laboratory,
where this work was completed, for hospitality.

\end{document}